\documentclass[sigconf,nonacm,numbers]{acmart}

\AtBeginDocument{%
  }

\setcopyright{acmlicensed}
\copyrightyear{2018}
\acmYear{2018}
\acmDOI{XXXXXXX.XXXXXXX}

\acmConference[Conference acronym 'XX]{Make sure to enter the correct
  conference title from your rights confirmation emai}{June 03--05,
  2018}{Woodstock, NY}
\acmISBN{978-1-4503-XXXX-X/18/06}

\usepackage{amsmath,amsfonts}
\usepackage{algorithmic}
\usepackage{graphicx}
\usepackage{url}
\usepackage{xcolor}
\usepackage{natbib}
\usepackage{xspace}
\usepackage{tcolorbox}
\usepackage{booktabs}
\usepackage{adjustbox}
\usepackage{hyperref}
\usepackage{multirow}
\usepackage{caption}
\usepackage{subcaption}
\usepackage[group-digits=integer,group-minimum-digits=4,group-separator={{,}},per-mode=symbol,detect-all]{siunitx}
\usepackage{colortbl}
\usepackage{enumitem}
\usepackage{array}
\usepackage{balance}

\newif\ifDraft

\Drafttrue

\usepackage[lambda,adversary,asymptotics,sets,landau,probability,operators]{cryptocode}

\makeatletter
\g@addto@macro{\UrlBreaks}{\UrlOrds}
\makeatother

\usepackage{cleveref}

\usepackage{seqsplit}
\newcommand{\ttt}[1]{%
  \begingroup
    \protect\renewcommand{\seqinsert}{\ifmmode\allowbreak\else\-\fi}%
    \protect\texttt{\protect\seqinsert{\protect\seqsplit{\small#1}}}%
  \endgroup
}
\newcommand{\tttscript}[1]{%
  \begingroup
    \protect\renewcommand{\seqinsert}{\ifmmode\allowbreak\else\-\fi}%
    \protect\texttt{\protect\seqinsert{\protect\seqsplit{\scriptsize#1}}}%
  \endgroup
}
\newcommand{\tttfoot}[1]{%
  \begingroup
    \protect\renewcommand{\seqinsert}{\ifmmode\allowbreak\else\-\fi}%
    \protect\texttt{\protect\seqinsert{\protect\seqsplit{\footnotesize#1}}}%
  \endgroup
}

\newcommand{\txtarget}{$T_{Target}$\xspace}
\newcommand{\txmev}{$T_{MEV}$\xspace}

\newcolumntype{L}[1]{>{\raggedright\let\newline\\\arraybackslash\hspace{0pt}}m{#1}}
\newcolumntype{C}[1]{>{\centering\let\newline\\\arraybackslash\hspace{0pt}}m{#1}}
\newcolumntype{R}[1]{>{\raggedleft\let\newline\\\arraybackslash\hspace{0pt}}m{#1}}

\begin{document}

\title[Rolling in the Shadows: Analyzing the Extraction of MEV Across Layer-2 Rollups]{Rolling in the Shadows: \\ Analyzing the Extraction of MEV Across Layer-2 Rollups}

\author{Christof Ferreira Torres}
\authornote{The author contributed to this work while affiliated with ETH Zurich.}
\orcid{0000-0001-6992-703X}
\affiliation{%
  \institution{Instituto Superior Técnico \& University of Lisbon \& INESC-ID}
  \city{Lisbon}
  \country{Portugal}
}
\email{christof.torres@tecnico.ulisboa.pt}

\author{Albin Mamuti}
\orcid{0000-0003-0796-7962}
\affiliation{%
  \institution{ETH Zurich}
  \city{Zurich}
  \country{Switzerland}
}
\email{amamuti@student.ethz.ch}

\author{Ben Weintraub}
\orcid{0000-0002-9527-5888}
\affiliation{%
  \institution{Northeastern University}
  \city{Boston}
   \state{Massachusetts}
  \country{USA}
}
\email{weintraub.b@northeastern.edu}

\author{Cristina Nita-Rotaru}
\orcid{0000-0002-9649-6789}
\affiliation{%
  \institution{Northeastern University}
  \city{Boston}
   \state{Massachusetts}
  \country{USA}
}
\email{c.nitarotaru@northeastern.edu}

\author{Shweta Shinde}
\orcid{0000-0003-0415-2960}
\affiliation{%
  \institution{ETH Zurich}
  \city{Zurich}
  \country{Switzerland}
}
\email{shweta.shinde@inf.ethz.ch}


\begin{abstract}
The emergence of decentralized finance has transformed asset trading on the blockchain, making traditional financial instruments more accessible while also introducing a series of exploitative economic practices known as  Maximal Extractable Value (MEV). Concurrently, decentralized finance has embraced rollup-based Layer-2 solutions to facilitate asset trading at reduced transaction costs compared to Layer-1 solutions such as Ethereum. However, rollups lack a public mempool like Ethereum, making the extraction of MEV more challenging.

In this paper, we investigate the prevalence and impact of MEV on Ethereum and prominent rollups such as Arbitrum, Optimism, and zkSync over a nearly three-year period. Our analysis encompasses various metrics including volume, profits, costs, competition, and response time to MEV opportunities. We discover that MEV is widespread on rollups, with trading volume comparable to Ethereum. We also find that, although MEV costs are lower on rollups, profits are also significantly lower compared to Ethereum.
Additionally, we examine the prevalence of sandwich attacks on rollups. While our findings did not detect any sandwiching activity on popular rollups, we did identify the potential for cross-layer sandwich attacks facilitated by transactions that are sent across rollups and Ethereum. Consequently, we propose and evaluate the feasibility of three novel attacks that exploit cross-layer transactions, revealing that attackers could have already earned approximately 2 million USD through cross-layer sandwich attacks.
\end{abstract}

%

\keywords{Ethereum; layer-2; rollups; MEV; cross-layer; sandwiching; DeFi}

\maketitle

\section{Introduction}

Blockchain markets have revolutionized the finance industry.
Entirely new classes of assets are now regularly traded at massive scales.
Partially responsible are smart contract technologies and the subsequent advent of decentralized finance (DeFi), which create markets that are transparent and freely accessible to anyone with an internet connection.
As of April 10th, 2024, over 95 billion USD worth of assets have been deposited across the major DeFi smart contracts platforms~\cite{defillama}, a number comparable to the 90 billion USD market cap of well-known traditional banks such as UBS~\cite{companiesmarketcap}, thus grounding the magnitude of DeFi's popularity.

However, with DeFi's important position within the global economy comes commensurate traffic and trade volume.
Ethereum, the most popular among DeFi ecosystem blockchains~\cite{defillama_chains}, has seen an increase in traffic thus increasing competition to push transactions through a throughput-limited system.
This competition for a limited resource has resulted in increased transaction fees.
Several solutions have been proposed for increasing Ethereum's throughput.
While some of these solutions are improvements in the Ethereum protocol (e.g., sharding \cite{dang2019towards,yu2020survey}), others operate as a layer of abstraction over the underlying Ethereum blockchain. One method that has seen support by the Ethereum community, is the outsourcing of transaction processing to external solutions.
These solutions can be characterized into four categories; side-chains (e.g., Polygon \cite{polygon}), generic bridges (e.g., Ronin \cite{ronin}, Wormhole \cite{wormhole}, etc.), payment or state channels (e.g., Raiden \cite{raiden}, Perun \cite{perun}, etc.), and rollups (e.g., Arbitrum \cite{arbitrum}, zkSync \cite{zksync}).
This class of solutions is often referred to as \emph{Layer-2} or \emph{L2}, whereas Ethereum is referred to as \emph{Layer-1} or \emph{L1}.
Among the four categories, rollups have emerged to be the most adopted solution \cite{l2beat}. This is due to their low transaction fees, high transaction throughput, and compatibility with Ethereum.
At the time of writing, a simple transaction on Ethereum to send cryptocurrency to another address costs about \num{1.47} USD, whereas the same transaction could be performed on any of the popular rollups for less than \num{0.01} USD \cite{l2fees}.

Rollups operate by allowing users to lock their assets on Ethereum, which grants them access to trade equivalent assets on a rollup chain.
Users can make trades on the rollup as frequently and to whatever degree their capital allows.
Rollups perform periodic state checkpoints with the Ethereum blockchain, which consist of batches of compressed rollup transactions stored in a data field of a traditional Ethereum transaction.
Each rollup has its own ordering mechanism, a service called a \textit{sequencer}.
In most cases, this is a centralized server operated by the rollup developer following a first-come, first-served strategy.
Because rollups follow a centralized approach towards transaction ordering, the transaction throughput can be much higher than in Ethereum.
Moreover, as multiple rollup transactions can be compressed into a single Ethereum transaction, the high fees associated with the Ethereum transaction can be amortized across all rollup transactions of the same batch.

Maximal Extractable Value (MEV) has become an essential part of Ethereum \cite{ferreira2021frontrunner,Qin_Zhou_Gervais_2022}. The idea behind MEV is to extract value (i.e., monetary profit) from transactions performed by other entities on the blockchain by influencing their order of execution.
Several strategies have been proposed, the most popular ones being arbitrage, liquidation, and sandwiching \cite{ferreira2021frontrunner,Qin_Zhou_Gervais_2022}.
MEV is a double-edged sword. While strategies such as arbitrage and liquidation help maintaining markets healthy across different DeFi protocols (e.g., liquidation of unhealthy loans, price balancing across exchanges, etc.), they can also have negative side effects. For example, they result in increased transaction fees as a result of gas price auctions \cite{flashboys} or wasteful transactions due to failed attempts to extract MEV. 
Some strategies, such as sandwiching, are even considered entirely destructive. Sandwiching occurs, for example, by placing a buy order right before a pending trade (i.e., frontrunning) and a sell order right after it (i.e., backrunning), thereby forming a ``sandwich'' and resulting in traders making less profit on their trades.

In contrast to Ethereum, rollups do not provide a mempool that publicly advertises pending transactions. 
Only sequencers can see transactions before they are finalized. 
Thus, normal users can only make use of information contained in the latest block to extract MEV. 
This means that arbitrages and liquidations are still possible, but that traditional sandwich attacks are not possible anymore since these require extractors to identify victim transactions upfront and then frontrun them before they are finalized. 
Hence, only sequencers have the power to mount such attacks.
However, sequencers are assumed to be trusted entities and, therefore, users assume that sandwich attacks are not an issue on rollups. 

In this paper, we shed light into how much MEV is being extracted on popular rollups such as Arbitrum \cite{arbitrum}, Optimism \cite{optimism}, and zkSync \cite{zksync} as these are the rollups with the highest market share in terms of total value locked at the time of writing \cite{l2beat}. We analyze how MEV extraction on these rollups compares to Ethereum in terms of volume, profit, and costs across a 32 month period. 
We also analyze the usage of flash loans, code reuse and competition among MEV extractors as well as the time extractors take to respond to MEV opportunities. 
Furthermore, we check for traditional sandwiching on rollups, confirming that so far sequencers behave as intended and do not perform any evident sandwiching.
Nonetheless, after carefully studying the way transactions are sent across
rollups and Ethereum, we propose and evaluate three novel cross-layer sandwich attacks that normal users can leverage to mount sandwich attacks on rollups. These attacks exploit transactions that perform trades on rollups but which are emitted via Ethereum. Our simulation using past
mainnet data
reveals that attackers could have made roughly 2 million USD profit.  
Finally, we also discuss the generalization of our attacks and potential countermeasures.

\noindent
\newline
\textbf{Ethical Considerations.}
For ethical reasons, we demonstrate the feasibility of our attacks by executing them only on the testnet, thereby targeting only our own victim transactions. 
On the mainnet, we only perform a local simulation of our attacks and do not actually carry out the attacks by broadcasting the transactions.
The attacks performed on the testnet did not rely on real victim transactions, but on victim transactions emitted by us. Our simulation on real victim transactions on the mainnet cannot have an impact anymore on these victims as their transactions are already finalized on L1 and L2, and cannot be exploited anymore.

\noindent
\newline
\textbf{Contributions.} We summarize our contributions as follows:
\begin{itemize}[leftmargin=*]
 \item We conduct the first large-scale measurement of MEV practices across Arbitrum, Optimism, and zkSync, and compare them to Ethereum over a period of nearly 3 years.
 \item We present novel insights in terms of volume, profits, costs, flash loans, code reuse, competition, and response time to MEV opportunities across Ethereum and rollups.
 \item We propose three novel cross-layer sandwich attacks and simulate them using real mainnet data from Arbitrum and Optimism. Our simulation shows that attackers could make approximately 2 million USD profit via cross-layer sandwich attacks. 
\end{itemize}

\section{Background}

In this section, we provide background on Ethereum, rollups, and Maximal Extractable Value.

\subsection{Ethereum}
Ethereum is a blockchain and cryptocurrency system.
The Ethereum blockchain consists of a series of ordered blocks, which themselves contain a sequence of ordered transactions.
Transactions can be simple exchanges of the main currency, Ether (ETH), or they can contain more complex programmatic logic called smart contracts.

Smart contracts are Turing-complete programs executed via the Ethereum Virtual Machine (EVM). They could execute infinitely if not for a particular design intervention called \textit{gas}.
Gas represents a fee that is charged for every instruction executed by a smart contract.
The relationship between gas and instructions is deterministic (e.g., an instruction that adds two numbers always costs three gas units~\cite{ethereum}).
Gas is paid in the form of ETH in proportion to the so-called gas price, which dictates how much ETH one gas unit costs.
To prevent unexpectedly expensive computations, every transaction is provided a gas limit, which means that the execution will stop once the limit has been reached.
When a transaction is aborted or reverted, external changes are rolled back but the gas, and the resulting ETH, is still spent.

Transaction ordering is at the discretion of the block builders, who construct blocks of pending transactions and then send them to proposers who validate them and send them out to the network to be added to the head of the blockchain.
Rational block builders order transactions based on the gas fees they will accrue when the block is validated, thereby giving priority to transaction that yield a higher income.
In effect, a user can get a transaction to appear higher in a block by increasing the gas price they are willing to pay. 

Smart contracts cannot communicate directly with the off-chain world (e.g., programs running outside the blockchain) as the EVM runs as an isolated synchronous deterministic environment. However, the EVM allows smart contracts to emit \emph{events}, which allow developers to store data during execution that is persistent and quickly accessible by applications running outside the blockchain.

\subsection{Rollups}

\begin{figure}
    \centering
    \includegraphics[width=\columnwidth]{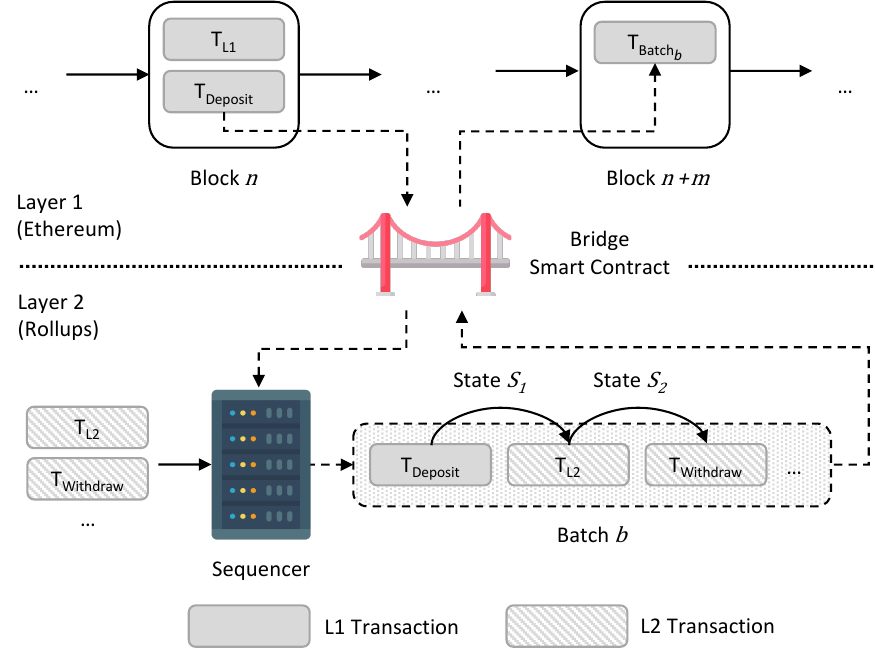}
    \caption{An overview on the interplay between Ethereum (Layer 1) and rollups (Layer 2).}
    \label{fig:rollups}
\end{figure}

Scalability and cost remain the salient issues of blockchain-based cryptocurrencies.
While numerous strategies have been proposed for increasing transaction throughput, from payment channel networks~\cite{poon2016bitcoin,raiden} to sharding~\cite{kokoris2018omniledger}, rollups have emerged as the favored solution for Ethereum.

A rollup derives its benefits by bringing the expensive computations off of Ethereum i.e., Layer-1 (L1), and uses L1 only for storage and state checkpointing.
This class of solutions and applications are called Layer-2 (L2) solutions, highlighting the fact that they are conceptually built on top of L1.
A user enters a rollup by depositing funds to an L1 smart contract, called a \emph{bridge}.
Bridges allow users to transfer assets between separate blockchains and/or layers (see \figureautorefname{} \ref{fig:rollups}).
In this case, it transfers assets from the L1 blockchain to the L2 rollup.
Deposited funds become spendable within the rollup but unspendable on L1.

Rollups can either be \emph{optimistic} or \emph{zero-knowledge}. Optimistic rollups assume that the sequencer is behaving correctly, but allows users to dispute any irregularities within a challenge period---it is a ``trust but verify'' model.
On the other hand, zero-knowledge rollups (ZK rollups) are untrusted from the start. 
Every ZK rollup must include a cryptographic validity proof to be considered valid.
When a user wishes to make a payment, they submit the payment directly to the \emph{sequencer}.
The sequencer accepts these transactions and organizes them into batches, which are analogous to blocks in L1.
It executes the transactions in an off-chain EVM to make sure they are valid and also to track the state changes that may result from contract execution.
\tableautorefname{} \ref{tab:rollup_comparison} provides a comparison between Ethereum and different rollup releases studied in this work. Currently there is only a single sequencer in each of the major rollup platforms and none of them provide a public mempool.
For most rollups the ordering of transactions follows a \emph{``first-come, first-served''} (FCFS) policy, whereas on Ethereum transaction ordering can be done arbitrarily at the discretion of the block builder.

Periodically, after assembling several batches, a sequencer compresses the batches and puts them into a raw data field---called \ttt{calldata}---in an L1 Ethereum transaction.
The sequencer then submits this well-formed Ethereum transaction to the L1 blockchain.
Its visibility and availability on L1 allow participants to look at the published transactions and dispute any inconsistencies.
Since the the sequencer is trusted to process these L2 transactions correctly, and should be able to do so using a standard EVM implementation, any deviations from correct behavior are assumed to be malicious.
For optimistic rollups, users can submit fraud proofs during the challenge period if the rollup is invalid for whatever reason. After the challenge period ends, the transactions can no longer be reversed and have achieved \emph{finality}.

Users can get their funds out of a rollup by submitting a request to the sequencer.
The sequencer includes a transaction in the rollup that \emph{burns} the user's funds on the rollup.
This means that the funds on the rollup are no longer spendable.
Meanwhile, a similar transaction, called a \emph{withdrawal proving transaction} is propagated to L1, which confirms that the L2 transaction did occur---this prevents a user from double spending both the L2 funds and their deposited L1 funds.
The last step is another L1 transaction called a \textit{withdrawal finalizing transaction} that confirms that the challenge period has ended, which ensures that the funds do not become spendable on L1 if they are disputed on L2.
Note that these extra assurances are not needed for ZK rollups, which results in simplified exit procedures.


\begin{table}[b]
    \centering
    \begin{adjustbox}{width=\columnwidth}
    \begin{tabular}{l l l l l}
        \toprule
        \textbf{Chain} & \textbf{Release} & \textbf{Rollup Type} & \textbf{Mempool} & \textbf{Ordering} \\
        \midrule
        Ethereum & -              &          - & Public  & (D) Gas Price \\
        Arbitrum & Classic        & Optimistic & Private & (C) FCFS \\
        Arbitrum & Nitro          & Optimistic & Private & (C) FCFS \\
        Optimism & Pre-Bedrock    & Optimistic & Private & (C) FCFS \\
        Optimism & Post-Bedrock   & Optimistic & Private & (C) Gas Price \\
        zkSync   & Era            & Zero-Knowledge & Private & (C) FCFS \\
        \bottomrule
        \multicolumn{5}{l}{(D): Decentralized, (C): Centralized} \\
        \\
    \end{tabular}
    \end{adjustbox}
    \caption{A comparison between Ethereum and rollups studied in this work (i.e., Arbitrum, Optimism, and zkSync).
}
    \label{tab:rollup_comparison}
\end{table}

\subsection{Maximal Extractable Value}

Transactions in Ethereum are typically publicly visible before they are officially committed to the blockchain via the mempool (i.e., pool of pending transactions).
Similar to traditional financial markets, Ethereum users have noticed that profit can sometimes be made off of these uncommitted transactions~\cite{flashboys_lewis}.
\citet{flashboys} first characterized this phenomenon, which we now call \emph{Maximal Extractable Value} (MEV).

In general, this value is extracted through a process of influencing transaction ordering.
There are many contexts in which this is applicable.
The Ethereum ecosystem allows for the creation of \textit{tokens} via smart contracts.
These act as individual currencies, with value independent of Ethereum's native currency, ETH.
A decentralized exchange (DEX) operates much like a traditional currency exchange.
A user requests to trade in some amount of one token for the equivalent value in another token.
The exchange rate can either come from an off-chain source, which is manually administered, or it can be modulated dynamically on-chain via so-called \emph{Automated Market Makers} (AMMs). 
The most popular DEX protocols across chains in terms of Total Value Locked (TVL) \cite{popular_dexes} are \emph{Uniswap} \cite{uniswap} (including forks such as \emph{SushiSwap} \cite{sushiswap}), \emph{Balancer} \cite{balancer}, and \emph{Curve} \cite{curve}. Apart from DEXes, lending platforms also play a prominent role in MEV extraction. Users can borrow assets from these platforms given that they provide another asset as collateral and payback the loan including some interest. At the time of writing, the most popular lending platforms across chains in terms of TVL \cite{popular_lending_platforms} are \emph{Aave} \cite{aave} and \emph{Compound} \cite{compound}.

\begin{figure}[t]
    \centering
    \includegraphics[width=\columnwidth]{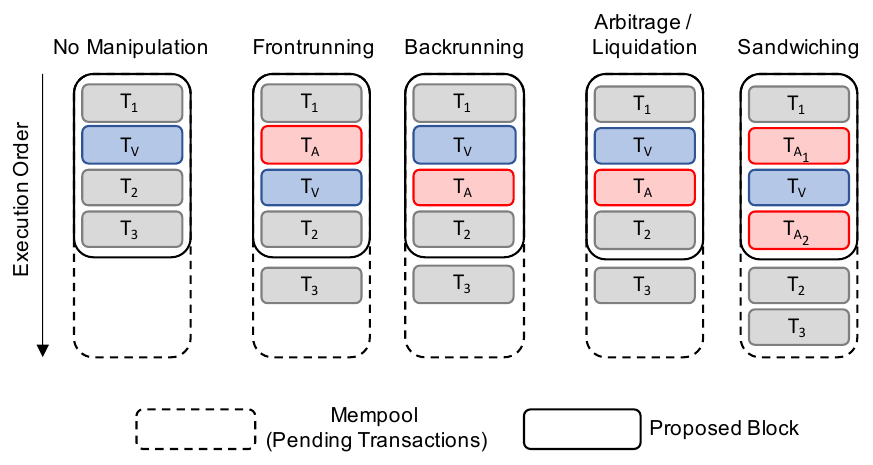}
    \caption{Examples of common MEV extraction strategies. Blue boxes depict victim transactions and red boxes depict MEV transactions for each type.}
    \label{fig:mev_taxonomy}
\end{figure}

There are two transaction ordering primitives that are leveraged to extract MEV: \emph{frontrunning} and \textit{backrunning} (see \figureautorefname{} \ref{fig:mev_taxonomy}).
Backrunning is when the MEV extractor ensures that their own transaction \txmev is only executed \emph{after} some target transaction \txtarget. Examples of backrunning include liquidating loans as quickly as possible or selling an asset that just had its price increase as the result of a large target transaction \txtarget.
The alternative ordering primitive is called frontrunning. 
Frontrunning is when an MEV extractor manipulates the ordering such that their own transaction \txmev precedes some target transaction \txtarget.
The motivation for such an action could be either to acquire some asset before its price changes as a result of a large victim transaction \txtarget, or to be the first to execute some contract (e.g., copying profitable transactions from other users~\cite{qin2023blockchain}).

There exist three popular MEV extraction techniques that exploit the aforementioned transaction ordering primitives: \emph{arbitrage}, \emph{liquidation}, and \emph{sandwiching} (see \figureautorefname{} \ref{fig:mev_taxonomy}). 
Arbitrage is the practice of concurrently selling and purchasing assets across different exchanges to capitalize on variations in market prices. Arbitrageurs engage in arbitrage by monitoring blockchain state changes. These changes can be monitored by either analyzing pending transactions via the mempool or by analyzing the state of the latest block. 
Liquidations enable users to purchase collateral at a discount when
repaying debt. The discount can either be fixed or determined via an auction. MEV extractors typically focus on liquidations with fixed discounts as these can be performed via a single transaction. Similar to arbitrage, liquidators either analyze the mempool or the state of the latest block to find opportunities.
Sandwiching is a classic trading strategy and well-known in traditional finance. 
It involves analyzing pending transactions and wrapping a victim's pending transaction $T_V$ within two adversarial transactions $T_{A_1}$ and $T_{A_2}$. Typically, $T_V$ aims to perform a large trade, whereas the sandwicher first frontruns $T_V$ by buying the same asset at a cheaper price and afterwards backruns $T_V$ to sell the purchased asset at a much higher price, thereby profiting from $T_V$ (see \Cref{fig:mev_taxonomy}).

Arbitrage and liquidation are generally viewed as ``good'' MEV as they play an essential role in fostering market health. 
For example, arbitrage helps DEXes in keeping their prices synchronized across other DEXes. 
However, sandwiching is considered ``bad'' MEV, since sandwichers manipulate the price that other traders get. 
But the dichotomy is not always so simple and even good MEV can have negative side effects. 
In particular, gas price auctions \cite{flashboys}---when multiple transactions are competing for preferential block position by increasing their gas price---results in increased transaction fees for all uses and produces block congestion as a result of the flood of MEV extractions. 
Flashbots aims to solve these issues via private mempools \cite{flashbots}. However, Flashbots only operates on Ethereum and rollups currently do not provide public mempools. 
Rollups also typically order transactions based on FCFS. Hence, sandwiching is theoretically only possible by the sequencer. 
Moreover, arbitrage and liquidation can be performed by anyone observing the latest state and optimizing transaction latency. 
Additionally, block congestion is still possible, since arbitrageurs and liquidators cannot observe competitors in real time as opposed to Ethereum.
\section{Detecting MEV Across Layers}\label{sec:method}

In this section, we describe our methodology on detecting arbitrage, liquidation, sandwiching, flash loans, MEV opportunities, and competition across Ethereum, Arbitrum, Optimism, and zkSync through the analysis of historical blockchain data.

\subsection{Detecting Arbitrage}

\begin{figure}
    \centering
    \includegraphics[width=\columnwidth]{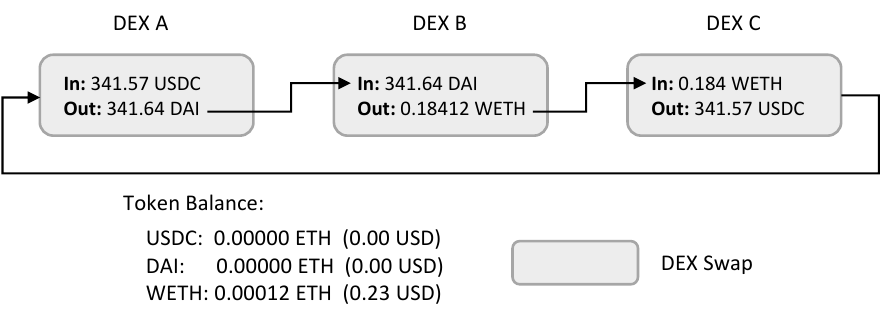}
    \caption{Example of our arbitrage detection across 3 swaps.}
    \label{fig:arbitrage_detection}
\end{figure}

We detect arbitrage based on heuristics proposed by previous works on detecting cyclic arbitrage in DEXes \cite{Qin_Zhou_Gervais_2022,flashbotinthepan,wang2022cyclic}. 
We start by scanning past blocks for token swap events emitted by DEXes such as Uniswap V2, Uniswap V3, Balancer V1, Balancer V2, and Curve (see \Cref{sec:events}). A swap event denotes a successful exchange of a token $A$ for another token $B$. In other words, $X$ amount of token $A$ going into the DEX smart contract and $Y$ amount of token $B$ going out of the DEX smart contract. DEXes may implement and call these events differently. For example, while Uniswap V2 and Uniswap V3 emit the same event called \emph{``Swap''}, their event topic hashes are different since they encode different values (e.g., \texttt{Swap(address,uint,uint,uint,uint,address)} for Uniswap V2 and \texttt{Swap(address,address,int256,int256,uint160,uint128,
int24)} for Uniswap V3). 
Once we retrieve all swap events, we extract information such as: token address in, token address out, token amount in, token amount out, and the address of the DEX that performed the swap. 
Afterwards, we group swaps by the transactions they are a part of and iterate through these swaps following the order in which they were emitted. 
Next, we link the swaps together, since arbitrages are essentially a chain of swaps. Hence, for each swap we check whether the token address out is equivalent to the token address in of the subsequent swap and whether the token amount out of the current swap is equivalent or higher than the token amount in of the subsequent swap (see \figureautorefname{} \ref{fig:arbitrage_detection}). Moreover, we also verify that the DEX address of the current swap is different from the DEX address of the subsequent swap. We find an arbitrage whenever we encounter a swap where the token address out is equivalent to the token address in of the very first swap of the current sequence, thus, forming a cycle. We restart the process for the remaining swap events to detect further arbitrages, since a single transaction may perform multiple arbitrages.

For every detected arbitrage, we keep track of the token balance of each token involved. We start with a balance of zero for every token and add or deduct the token amount depending on whether the token is going into the DEX (i.e., adding token amount) or going out of the DEX (i.e., deducting token amount). We then compute the cost of each arbitrage by converting all balances to Ether using CoinGecko's public API \cite{coingecko_api} and assigning negative balances to costs and positive balances to gains. The profit of the arbitrage is computed by simply deducting the costs from the gains. However, this does not return the final profit of the arbitrage transaction as the transaction fees are missing and a transaction may contain multiple arbitrages. Thus, we compute the final profit by first summing up all the profits of all the arbitrages involved in the transaction and then deducting the transaction fees. In case of Ethereum, we also check whether the transaction was part of a Flashbots bundle using Flashbots' public API \cite{flashbots_api}. If the transaction is part of a bundle, we deduct the coinbase transfer value associated to the transaction. The coinbase transfer is another way (besides traditional transaction fees) for MEV extractors to transfer funds to block producers. Hence, MEV extractors use coinbase transfers for bribing (i.e., incentivizing) block producers to prioritize the inclusion of their transaction in the next block.

\subsection{Detecting Liquidation}

We detect liquidations by scanning past blocks for specific liquidation events emitted by smart contracts of popular DeFi lending protocols. In this work, we analyze liquidations performed by smart contracts that follow \emph{Aave}'s \cite{ aave_flash_loan} and \emph{Compound}'s \cite{compound_liquidations} lending protocol implementations.
In detail, our detection script searches for Aave's V1, V2, and V3 \emph{``LiquidationCall''} event topic hashes and Compounds’s V2 \emph{``LiquidateBorrow''} event topic hash (see Appendix \ref{sec:events}). These events are triggered whenever a loan has been successfully liquidated. For each event we extract information such as liquidator, liquidated user, liquidated debt, received collateral, etc. 

Loans that follow Compound's protocol do not directly return the received collateral. 
Instead, users receive the collateral in form of ``cTokens'', which they can leave in Compound to gain interest, or redeem for the actual collateral. We therefore also scan the block for emitted \emph{``Redeem''} events, which allows us to obtain the actual collateral, including its actual amount.
We compute the profit for each liquidation by deducting the cost, which is the value of the debt token, from the gain which is the value of the collateral. Similar to the arbitrage detection, we convert the value of both, debt and collateral, into its equivalent
amount in Ether using CoinGecko’s public API \cite{coingecko_api}. 
MEV extractors might liquidate multiple loans within the same transaction. 
The final profit is computed by summing up all the profits of each liquidation and deducting the transaction costs as well as any tips that the
MEV extractor may have paid to the block proposer via coinbase transfers (the latter only applies if the MEV extractor performed the liquidation on Ethereum).

\subsection{Detecting Sandwiching}


We base our sandwich detection on a combination of heuristics proposed by previous works \cite{ferreira2021frontrunner,Qin_Zhou_Gervais_2022}. We begin by scanning past blocks for token transfer events. 
For Ethereum, we analyze the events per block, whereas for the rollups we scan token events across multiple blocks (i.e., batches) using a sliding window of 100 blocks with a distance of one block apart. 
Due to the small size of L2 batches, it is more likely that sandwiches will be spread across batches. 
Theoretically this could also be the case for Ethereum, however previous works \cite{ferreira2021frontrunner,Qin_Zhou_Gervais_2022} have observed that analyzing sandwiches per block is sufficient in practice. 

For each detected transfer event we aim to find its counterpart, namely a transfer event that was emitted afterwards via a different transaction by the same token smart contract but which transfers the tokens in the opposite direction. Hence, we search for transactions $T_{A_1}$ and $T_{A_2}$, such that $T_{A_2}$ has a larger transaction index value than $T_{A_1}$ and where the sender of $T_{A_1}$ is equivalent to the receiver of $T_{A_2}$ and the receiver of $T_{A_1}$ is equivalent to the sender of $T_{A_2}$. 
Moreover, the amount transferred in $T_{A_2}$ has to be either the same or smaller than the amount transferred in $T_{A_1}$. 
The same applies for rollups, except that we verify the order by first checking batch height and then within the same batch the transaction index.
Once such a pair has been identified, we search for one or multiple victim transactions (i.e., $T_V$), by checking if there are any token transfer events emitted by the same token in between the identified attacker events, which have the same sender as $T_{A_1}$ but a different receiver.
The sender for $T_{A_1}$ and $T_{V}$ are expected the same since it is the address of the same DEX, however the receiver must be different since it is the address of the actual attacker/victim.


\subsection{Detecting Flash Loans}

Flash loans are risk-free loans which enable users to borrow a large amount of assets without being required to provide any collateral as a security \cite{wang2021towards}. Flash loans exploit the atomicity of blockchain transactions. Users are required to pay back the loan plus some interest within the same transaction. Otherwise the entire transaction is reverted, meaning that the transaction has no effect and the borrowed assets are returned back to the flash loan provider. There exist several DeFi protocols that provide flash loans. 

In this work, we analyze and compare two of the largest protocols: \emph{Aave} \cite{aave_flash_loan} and \emph{Balancer} \cite{balancer_flash_loan} 
These two protocols are deployed on Ethereum, Arbitrum, Optimism, and zkSync.
Flash loans enable MEV extractors to take out large quantities of MEV from protocols without requiring them to own upfront any asserts in those protocols.
Extractors are only required to own enough funds to pay back the
interest and the execution costs of the loan. However, due to flash loans being bound to the execution of a single transaction, they can only be used to extract liquidations and arbitrages but not sandwiches as these span across multiple transactions. 

For every detected arbitrage and liquidation, we scout for specific events that are emitted by flash loan providers whenever a flash loan was successful (i.e., the loan has been borrowed and paid back with interest). 
In more detail, for every transaction we check for the \emph{``FlashLoan''} event topic hashes emitted by Aave V1, Aave V2, Aave V3, and Balancer (see Appendix \ref{sec:events}).
We parse the event data for each protocol individually and extract information such as borrowed token, borrowed amount, interest fee, etc.

\subsection{Detecting Opportunities}

\begin{figure}
    \centering
    \includegraphics[width=0.9\columnwidth]{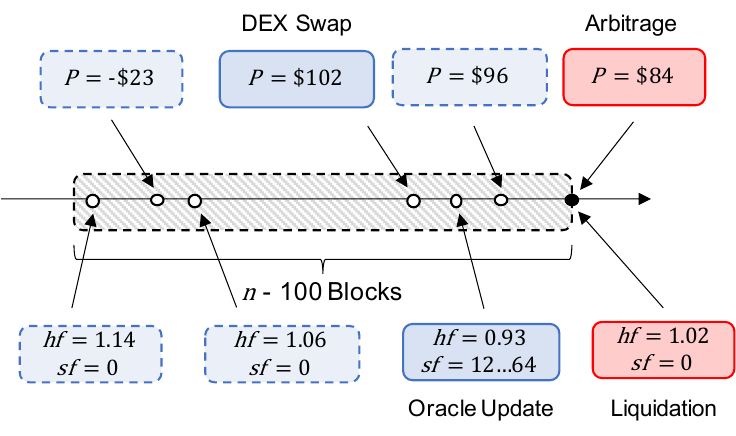}
    \caption{Example of our opportunity detection methodology. $P$ means profit and a change from a negative $P$ to a positive $P$ indicates an opportunity for arbitrage. An $hf$ below 1.0 or an $sf$ above zero  indicates an opportunity for liquidation. Boxes on top show arbitrage, boxes below show liquidation.}
    \label{fig:opportunity_detection}
\end{figure}

An opportunity is a transaction that triggers a state change which enables the extraction of MEV. For example, for an arbitrage to be possible, there needs to be a price difference for the same asset across two different exchanges.
This is common for DEXes as the price of an asset is determined by their local view on supply and demand. As a result, DEXes are not aware of price changes across other DEXes. Thus, a large trade on one DEX might create a large price difference for a particular asset, which can then be exploited by trading the asset across different exchanges.

Detecting transactions that triggered sandwiches is straightforward, since by definition sandwiches encompass the opportunity transaction (i.e., the victim transaction $T_{V}$).
The challenge resides in detecting transactions that triggered arbitrages and liquidations as these are (most often) decoupled from the actual transactions performing the MEV extraction.


\subsubsection{Detecting Arbitrage Opportunities}

Arbitrage opportunities are typically triggered via transactions performing large trades which cause price fluctuations for an asset on a DEX. 
Therefore, for every arbitrage that we detect, we retrieve all previous DEX swap events that are up to 100 blocks prior to the detected arbitrage (see \figureautorefname{} \ref{fig:opportunity_detection}). Our preliminary results have shown that 100 blocks are enough to capture most opportunities (see \sectionautorefname{} \ref{sec:opportunities}). 
Next, we filter out all swap events which do not originate from a DEX that was involved in the arbitrage. 
For each of the remaining swap events we simulate the arbitrage using the events' block number. We start with the largest block number (i.e., closest block to the arbitrage) and iterate towards the smallest block number (i.e., farthest block from the arbitrage). We stop iterating when we find a block where the simulation of the arbitrage returns a negative profit $P$. This indicates that the arbitrage is not profitable anymore, hence the swap transaction that triggered the arbitrage must be included within the last block where the arbitrage was profitable.

\subsubsection{Detecting Liquidation Opportunities}

A loan typically opens up for liquidation when the collateral that was provided by a user as a security for a borrowed asset loses significant value. In case of Aave, this is observable via the \emph{health factor} (\textit{hf}) which can be obtained by calling the smart contract function \texttt{getUserAccountData
(address user)} \cite{health_factor_aave}. If the health factor of a loan is below $1.0$, then the collateral lost a significant amount of value and the loan is open for liquidation. In case of Compound, this is observable via the \emph{shortfall} (\textit{sf}) which can be obtained by calling the smart contract function \texttt{getAccountLiquidity(address account)} \cite{shortfall_compound}. Shortfall is a positive integer and defines the amount by which a loan exceeds the required collateral. Hence, shortfall should always be zero. A loan that has a shortfall higher than 0 is liquidable. To know the current value of a collateral, protocols such as Aave and Compound rely on price oracles. Most instances of these two protocols make use of off-chain data that is available via Chainlink's price oracles \cite{chainlink_price_oracle}. Thus, a loan typically opens up for liquidation as a result of a price update via a Chainlink oracle.

Similar to our arbitrage opportunity detection, we detect liquidation opportunities by first retrieving all previous Chainlink \emph{``AnswerUpdated''} events (see Appendix \ref{sec:events}) that are up to 100 blocks away from the detected liquidation (see \figureautorefname{} \ref{fig:opportunity_detection}). For each of the oracle update events we obtain the block number and retrieve either the \textit{hf} or the \textit{sf} for Aave or Compound, respectively. Again, we start with the largest block number (i.e., closest block to the liquidation) and iterate towards the smallest block number (i.e., farthest block from the liquidation). We stop iterating when we find a block where either \textit{hf} is larger or equal to $1.0$ or \textit{sf} is equal to $0$. This indicates that the liquidation is not open anymore. Hence, the oracle update transaction that triggered the liquidation must be included in the previous block where \textit{hf} is smaller than $1.0$ or \textit{sf} is larger than $0$.

\subsection{Detecting Competition}

We define competition as two or more MEV extractors targeting the same MEV opportunity. Hence, we measure competition by analyzing whether two or more transactions that have been previously identified to extract the same type of MEV, have also been identified to target the same opportunity transaction as defined by our methodology on detecting MEV opportunities.


\subsection{Limitations}

Our methodology currently does not capture all MEV as it searches for hard-coded events triggered by a specific set of DeFi protocols, thus we will miss MEV emitted by other protocols. 
Hence, our results should be considered as a lower bound on the volume of extraction as well as in terms of comparison between DeFi protocols. 
Moreover, our competition measurement does not capture MEV transactions that did not make it on-chain. 
Similar to prior works \cite{Qin_Zhou_Gervais_2022,McLaughlinKV23,ferreira2021frontrunner,flashbotinthepan}, which measured MEV on L1, our price calculation relies on CoinGecko's public API \cite{coingecko_api} to identify the value of a token at a given time. 
As a result, the profit calculation might not precisely reflect the actual profit that the extractor made on that day. 
Moreover, CoinGecko only tracks popular tokens, thus, for less popular tokens we are unable to calculate the profits made by the extractor.
\section{Analyzing MEV Across Layers}

\begin{table}[t]
    \centering
    \begin{adjustbox}{width=\columnwidth}    
    \begin{tabular}{l c r p{0.11cm} r}
        \toprule
        \textbf{Chain} & \textbf{Time Period} & \multicolumn{3}{c}{\textbf{Block Range}} \\
        \midrule
        Ethereum & Jan. 1, 2021 - Aug. 31, 2023 & 11,565,019 & - & 18,037,987 \\
        Arbitrum & Mar. 28, 2021 - Aug. 31, 2023 & 0 & - & 126,855,340 \\
        Optimism & Jan. 14, 2021 - Aug. 31, 2023 & 0 & - & 108,963,811 \\
        zkSync   & Feb. 14, 2023 - Aug. 31, 2023 & 0 & - & 12,689,375 \\
        \bottomrule
    \end{tabular}
    \end{adjustbox}
    \caption{Data collection time frame for each chain.}
    \label{tab:data_collection}
\end{table}

In this section, we analyze MEV extracted between January 1st, 2021 and August 31st, 2023 (2 years and 8 months) across Ethereum, Arbitrum, Optimism, and zkSync (see \tableautorefname{} \ref{tab:data_collection} for block ranges)
using the methodology discussed in \Cref{sec:method}.

\subsection{Volume}\label{sec:volume}

\begin{figure*}
    \centering
    \includegraphics[width=\textwidth]{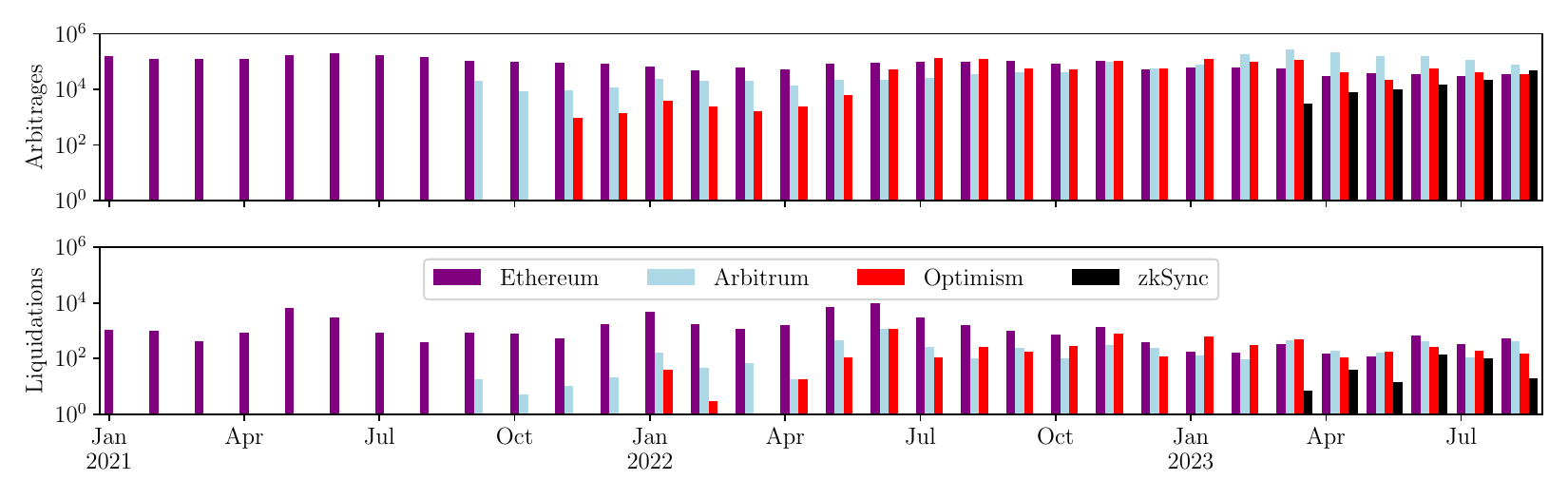}
    \caption{Number of detected arbitrages and liquidations per month on Ethereum, Arbitrum, Optimism, and zkSync.}
    \label{fig:arb-liq-volume}
\end{figure*}

For each of the rollups, we see gradual but significant increases in the number of liquidations and arbitrages over the course of our measurement period.
In the bottom plot of \Cref{fig:arb-liq-volume} we see a lot of variability in the number of liquidations on Ethereum, vacillating between 498 and 13,121 for each four month period. For Arbitrum and Optimism, we see rapid increase in the early months after the technology was deployed. Arbitrum then stabilizes between 309 and 665 liquidations per month. Optimism appears to decrease after its peak of 1,307 liquidations in the summer of 2022.
Interestingly, there is a decline in the number of liquidations on Ethereum between June 2022 and June 2023.
This could be a result of a smaller number of loans or broader economic trends causing the collaterals to maintain their value with higher probability.

We see similar trends in the top plot of \Cref{fig:arb-liq-volume}.
The salient difference between the two plots, though, is that all three of the rollups eventually surpass Ethereum in the number of detected arbitrages.
In the month of April 2023, Arbitrum had \num{7.2}$\times$ more arbitrages than Ethereum, although the difference narrows in the subsequent months.
Again, this is likely related to broader economic trends and the increasing usage of rollups as the transaction fees tend to be much lower.
After opening up to developers in February 2023, the first MEV was detected around March 25, 2023 on zkSync, highlighting the rapid adoption of MEV extractors even on ZK-based rollups.
Finally, we found no instances of sandwiching on any of the rollup platforms suggesting that the sequencers are not performing sandwich trades.
We provide additional granularity in 
 \Cref{tab:arb_lib_count}  in 
Appendix \ref{sec:appendix_c}.
However, in \Cref{sec:cross}, we show that cross-layer sandwiching is a viable threat vector to traders on rollups. Cross-layer sandwich attacks can be mounted by normal users and do not require control over pending transactions.




\subsection{Profits}

\begin{table}
    \centering
    \begin{adjustbox}{width=\columnwidth}
    \begin{tabular}{llrrrr}
        \toprule
        \textbf{Strategy} & \textbf{Profit} & \textbf{Ethereum} & \textbf{Arbitrum} & \textbf{Optimism} & \textbf{zkSync}\\
        \midrule
        \multirow{5}{*}{Arbitrage} & Total & \num{209981692.18} & \num{17419113.84} & \num{2514046.36} & \num{640243.36}\\
        & Max & \num{25614599.91} & \num{1913418.44} & \num{79927.65} & \num{23891.42}\\
        & $\text{P}_{90}$  & \num{87.51} & \num{6.55} & \num{1.94} & \num{7.52}\\
        & Mean & \num{74.12} & \num{10.14} & \num{2.21} & \num{6.12}\\
        & Median & \num{4.80} & \num{0.27} & \num{0.23} & \num{0.83}\\
        & Min & \num{-15828.08} & \num{-1659.02} & \num{-15963.71} & \num{-21.39}\\
        \midrule
        \multirow{5}{*}{Liquidation} & Total & \num{232498892.47} & \num{856503.70} & \num{1308358.38} & \num{10733.13}\\
        & Max & \num{2675295.15} & \num{71224.66} & \num{288573.28} & \num{4014.54}\\
        & $\text{P}_{90}$ & \num{4814.07} & \num{225.64} & \num{74.82} & \num{60.10}\\
        & Mean & \num{5323.75} & \num{206.44} & \num{296.08} & \num{42.59}\\
        & Median & \num{116.91} & \num{1.60} & \num{0.67} & \num{4.98}\\
        & Min & \num{-127586.84} & \num{-821.46} & \num{-2422.01} & \num{-1682.03}\\
        \midrule
        & Total & \num{442480584.65} & \num{18275617.54} & \num{3822404.73} & \num{650976.49} \\
        \bottomrule
    \end{tabular}
    \end{adjustbox}
    \caption{Profits in USD made from MEV across each chain.}
    \label{tab:profits}
\end{table}

In \Cref{tab:profits}, we see that Ethereum is still the platform with the highest profits, both cumulatively and on average.
In fact, the cumulative profits from MEV on Ethereum are close to 10$\times$ the profits on the other three platforms combined for both arbitrage and liquidation.
This is likely because Ethereum is simply more popular and thus has more pools and more opportunities and more liquidity.
Additionally, we note that the mean values are much higher than the median values while the maximum values are much higher than the P90 values for all platforms---this indicates that a small percentage of transactions account for a disproportionate fraction of the total profits.
So while by definition, 90\% of transactions are below the P90 in terms of profits, those that are above the 90th percentile are way above.
Among the rollups, Aribtrum has significantly greater arbitrage profits than either Optimism or zkSync. 
While for liquidations, Optimism has slightly higher profits.
We suspect that the arbitrage profits on Arbitrum are higher due to users performing trades with higher amounts as compared to Optimism. We also suspect that liquidation profits are higher on Optimism due to higher collaterals.



\subsection{Costs}

\begin{figure}
    \centering
    \includegraphics[width=\columnwidth]{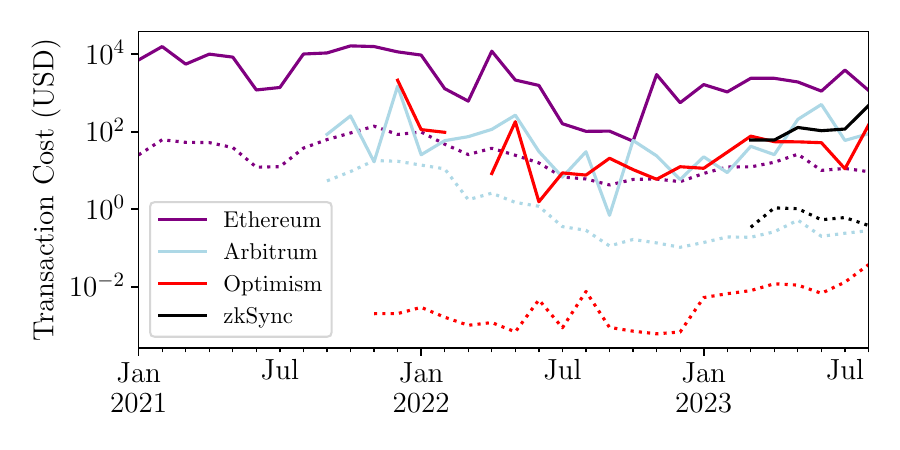}
    \caption{Median monthly transaction cost in USD for extracted MEV on each chain. Dotted lines depict arbitrages, while solid lines depict liquidations.}
    \label{fig:cost}
\end{figure}

In \Cref{fig:cost}, we present the median monthly transaction costs in USD over our measurement period for each platform.
The first insight we can observe is that liquidation transactions are significantly more expensive on all platforms.
The one exception being in November 2021, when both arbitrages and liquidations on Arbitrum cost the same.
For both arbitrage and liquidation, respectively, we see that transactions on Ethereum are more expensive than any of the rollups.
This suggests that these rollup platforms are achieving one aspect of their goal: making transactions cheaper.


%

\subsection{Opportunities}
\label{sec:opportunities}

\begin{figure}
    \centering    
    \includegraphics[width=1.0\columnwidth]{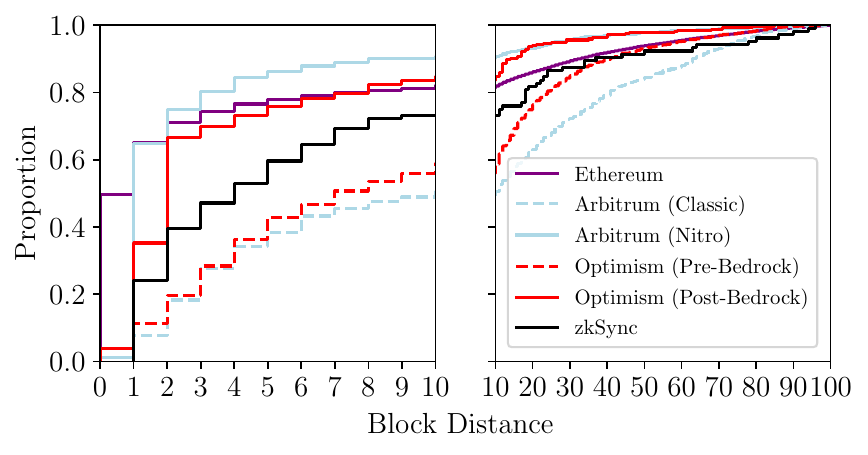}
    \caption{Block distance between extraction and opportunity.}
    \label{fig:opportunity_block_distance}
\end{figure}

\figureautorefname{} \ref{fig:opportunity_block_distance} depicts the cumulative distribution of the block distance between extracted MEV and MEV opportunities for up to a distance of 100 blocks. We see that for Ethereum, over 50\% of the MEV extractions occurred within the same block as the opportunity (i.e., block distance of $0$). Interestingly, for zkSync and former releases of Arbitrum and Optimism (i.e., Classic and Pre-Bedrock, respectively) we find that the smallest block distance is 1 (i.e., the MEV extraction never occurred in the same block as the opportunity). 
However, for newer releases of Arbitrum and Optimism (i.e., Nitro and Post-Bedrock, respectively) we do find a small amount of MEV extractions that occur within the same block. This is an interesting result since newer releases of Arbitrum and Optimism still do not have a public mempool. Yet, some extractors were able to insert their transactions within the same block as the opportunity. For instance, according to our dataset, address \href{https://optimistic.etherscan.io/address/0x4b94062fedd0dae7b623ad5e56c026a7db5e3021}{0x4b9406...5e3021} performed a total of six successful liquidations on Optimism, where all liquidations were performed within the same block as the opportunity, which hints towards a systematic approach. One example is block \num{108560468}, which contains 18 transactions and the extractor's transaction was included at position 11 whereas the opportunity is included at position 4. The block was built on August 22, 2023 i.e., towards the end of our measurement period.



\subsection{Competition}

For Ethereum, the largest number of MEV competitors that targeted the same arbitrage opportunity and were successful was 14.
Whereas for Arbitrum, Optimism, and zkSync it was 15, 9, and 7, respectively. 
Similarly, the largest number of MEV competitors that targeted the same liquidation opportunity and were successful was 10, 5, 5, and 0, for Ethereum, Arbitrum, Optimism, and zkSync, respectively.
This confirms that there is typically more competition on Ethereum than on rollups regarding MEV extraction. Moreover, we used the public APIs of Etherscan \cite{etherscan_api}, Arbiscan \cite{arbiscan_api}, Optimistic Etherscan \cite{optimistic_etherscan_api}, and zkSync's Block Explorer \cite{zksync_explorer_api} to obtain a list of transactions performed by each competitor and computed the percentage of reverted transactions across chains. For arbitrage, Ethereum has overall a rate of 8\% of reverted transactions among competitors, whereas Arbitrum, Optimism, and zkSync have 39\%, 29\%, and 38\%, respectively. A similar pattern is seen for liquidation, where Ethereum has overall a rate of 17\% of reverted transactions, whereas Arbitrum, Optimism, and zkSync have 29\%, 27\%, and 0\% (due to no liquidation competition detected), respectively. This highlights the advantage of proposer-builder separation (PBS) \cite{HeimbachKTW23} on Ethereum, where builders control which MEV transactions are forwarded to proposers. This allows them to forward only transactions that successfully extract MEV, thus reducing the amount of reverted transactions.



\subsection{Flash Loans}

We compare flash loan usage between rollups and Ethereum in \tableautorefname{} \ref{tab:flash_loans}. As expected, we observe that flash loans are not widely used for arbitrage, but are often used for liquidations.
Moreover, we see that flash loans are more popular on rollups than on Ethereum. 
For example, for liquidations, only 5.64\% of the MEV extractors on Ethereum used a flash loan, whereas on Arbitrum, Optimism, and zkSync they are used at rates of 48.88\%, 58.95\%, and 22.99\%, respectively. 
In addition, we find that Aave is used more often for liquidations on Arbitrum and Optimism, while Balancer is used more often on zkSync.
Interestingly, for arbitrages, we find that Balancer is the most used flash loan provider across all platforms.

\begin{table}
    \centering
    \begin{adjustbox}{width=\columnwidth}    
    \begin{tabular}{l l r r r r}
    \toprule
    \textbf{Strategy} & \textbf{Protocol} & \textbf{Ethereum} & \textbf{Arbitrum} & \textbf{Optimism} & \textbf{zkSync} \\
    \midrule
    \multirow{2}{*}{Arbitrage} &
    Aave & 
    0.08\%
    & 
    0.21\%
    & 
    0.03\%
    & 
    0.00\%
    \\
    & Balancer & 
    0.16\% 
    & 
    6.50\%
    & 
    13.20\%
    & 
    0.00\% 
    \\
    \midrule
    \multirow{2}{*}{Liquidation} & 
    Aave & 
    4.82\% 
    & 
    39.78\%
    & 
    44.38\% 
    & 
    2.37\%
    \\
    & Balancer & 
    0.82\% 
    & 
    9.17\% 
    & 
    14.57\%
    & 
    20.62\%
    \\
    \bottomrule
    \end{tabular}
    \end{adjustbox}
    \caption{Number of flash loans across rollups and Ethereum.}
    \label{tab:flash_loans}
\end{table}

\subsection{Code Reuse} 

\begin{table}[b]
    \centering
    \begin{adjustbox}{width=\columnwidth}    
    \begin{tabular}{l c c c}
        \toprule
        \textbf{Strategy} & \textbf{Ethereum} & \textbf{Arbitrum} & \textbf{Optimism} \\
        \midrule
        \multirow{6}{*}{\textbf{Arbitrage}} & \href{https://etherscan.io/address/0xF51Fe29A7DbD6E355f735B4fd91140416C3b677D}{\normalsize{\texttt{0xF51Fe2...3b677D}}} &
        \href{https://arbiscan.io/address/0x570CA28f482daC48Ee279764f25e731C6B16edf8}{\normalsize{\texttt{0x570CA2...16edf8}}} & - \\
        & \href{https://etherscan.io/address/0x1A8f43e01B78979EB4Ef7feBEC60F32c9A72f58E}{\normalsize{\texttt{0x1A8f43...72f58E}}} &
        \href{https://arbiscan.io/address/0x1A8f43e01B78979EB4Ef7feBEC60F32c9A72f58E}{\normalsize{\texttt{0x1A8f43...72f58E}}} & 
        \href{https://optimistic.etherscan.io/address/0x1A8f43e01B78979EB4Ef7feBEC60F32c9A72f58E}{\normalsize{\texttt{0x1A8f43...72f58E}}} \\
        & \href{https://etherscan.io/address/0x51C7D6760f80FFF75F27dDc52f2F049EcFf0C9c9}{\normalsize{\texttt{0x51C7D6...f0C9c9}}} &
        \href{https://arbiscan.io/address/0x568FBEFAeDd9aeEB7c981A4059BC5cF5C085D328}{\normalsize{\texttt{0x568FBE...85D328}}} & - \\
        & \href{https://etherscan.io/address/0x143647E484e0700D939b31Aa7C4A94A8155991b5}{\normalsize{\texttt{0x143647...5991b5}}} &
        \href{https://arbiscan.io/address/0x8076E6F8D00741d28Ef1201dAF12C57B35A94AFC}{\normalsize{\texttt{0x8076E6...A94AFC}}} & - \\
        & \href{https://etherscan.io/address/0xfa6d80aFE0ECf4EA916bBE4871dd6C2A1e1EBF90}{\normalsize{\texttt{0xfa6d80...1EBF90}}} &
        \href{https://arbiscan.io/address/0xa062B22b6E5b2c71d949bb6FdA3afBb1eCB2B84d}{\normalsize{\texttt{0xa062B2...B2B84d}}} & 
        \href{https://optimistic.etherscan.io/address/0xB4a513FD027B06a0E50F8fF40B98f51347Cac090}{\normalsize{\texttt{0xB4a513...Cac090}}} \\
        & \href{https://etherscan.io/address/0x2a2Cdd51b3C4a761BeE4A30Ea26C5a60dAe80C63}{\normalsize{\texttt{0x2a2Cdd...e80C63}}} & - & 
        \href{https://optimistic.etherscan.io/address/0xE8Ab13a84357F5D2C89e9E7D6521d83D43FBE6c7}{\normalsize{\texttt{0xE8Ab13...FBE6c7}}} \\
        \midrule
        \multirow{2}{*}{\textbf{Liquidation}} & \href{https://etherscan.io/address/0x81c3B160aE5cDAB6F82d16Df027A387aF14E4aEB}{\normalsize{\texttt{0x81c3B1...4E4aEB}}} &
        \href{https://arbiscan.io/address/0x8FEe8E0b9A3C2d5245382327F98e96B5C74f745D}{\normalsize{\texttt{0x8FEe8E...4f745D}}} & 
        \href{https://optimistic.etherscan.io/address/0x4540f200A064bb7a05D9D8C770BFDadB0e3d0598}{\normalsize{\texttt{0x4540f2...3d0598}}} \\
        & \href{https://etherscan.io/address/0x373f2b9f125F71cA46e2642A36319Eb24f8Ac495}{\normalsize{\texttt{0x373f2b...8Ac495}}} &
        \href{https://arbiscan.io/address/0x71a073cDF8364e6f3FB5714779c129639E7d2D82}{\normalsize{\texttt{0x71a073...7d2D82}}} & 
        \href{https://optimistic.etherscan.io/address/0x373f2b9f125F71cA46e2642A36319Eb24f8Ac495}{\normalsize{\texttt{0x373f2b...8Ac495}}} \\
        \bottomrule
    \end{tabular}
    \end{adjustbox}
    \caption{MEV bots with identical bytecode across chains.}
    \label{tab:identical_bots}
\end{table}


We explore whether MEV extractors reuse their smart contract bytecode across rollups and Ethereum. Both Arbitrum and Optimism are EVM compatible, meaning that developers can simply redeploy the same smart contracts on Arbitrum and Optimism which they previously deployed on Ethereum, without having to recompile them. Smart contracts on zkSync on the other hand are not EVM compatible.
Hence, developers are required to use a dedicated compiler~\cite{zkcompiler} to translate the EVM bytecode into a format that is executable on zkSync \cite{zkevm}.
To that end, we only download the runtime bytecode of smart contract addresses that are the destination (i.e., ``to'' field) of transactions that we previously identified performing either arbitrage or liquidation on Ethereum, Arbitrum, and Optimism. We ignore any bytecode that is verified on Etherscan and that contains the opcode \texttt{DELEGATECALL}. The intuition behind the former is that bots will not have their source code verified on Etherscan, whereas the latter prevents counting proxy smart contracts as bots \cite{smart_contract_proxies}.
Finally, we remove every \texttt{PUSH} opcode and its associated data from the bytecode including any hard-coded strings and metadata that are appended to the end of the bytecode \cite{solidity_metadata} as this data would be case-specific. 

For arbitrages, we identified \num{2160}, \num{873}, and \num{454} bots on Ethereum, Arbitrum, and Optimism, respectively.
While for liquidations, we found \num{484}, \num{101}, and \num{87} bots on Ethereum, Arbitrum, and Optimism, respectively. 
Overall, it seems that more users engage in arbitrage than liquidation, on both rollups and Ethereum. 
We checked whether the exact same bytecode is deployed across the same chain. 
For arbitrage, we found 27 clusters of identical bytecode on Ethereum, with the largest cluster containing 318 identical contracts. 
For Arbitrum and Optimism we found 16 different clusters, with the largest cluster containing 11 contracts on Arbitrum and 16 on Optimism. For liquidation, we also found clusters of identical contracts, although less as compared to arbitrage. We found six clusters on Ethereum with the largest one containing 10 identical contracts and two clusters on Arbitrum with the largest one containing three identical contracts. 
We did not find identical contracts on Optimism.
Finally, we checked for identical bytecode across different chains. 
We found three arbitrage bots and two liquidation bots with identical bytecode deployed across Ethereum, Arbitrum, and Optimism (see \tableautorefname{} \ref{tab:identical_bots}).

\section{Cross-Layer Sandwich Attacks}\label{sec:cross}

\begin{figure*}
    \centering
    \includegraphics[width=0.9\textwidth]{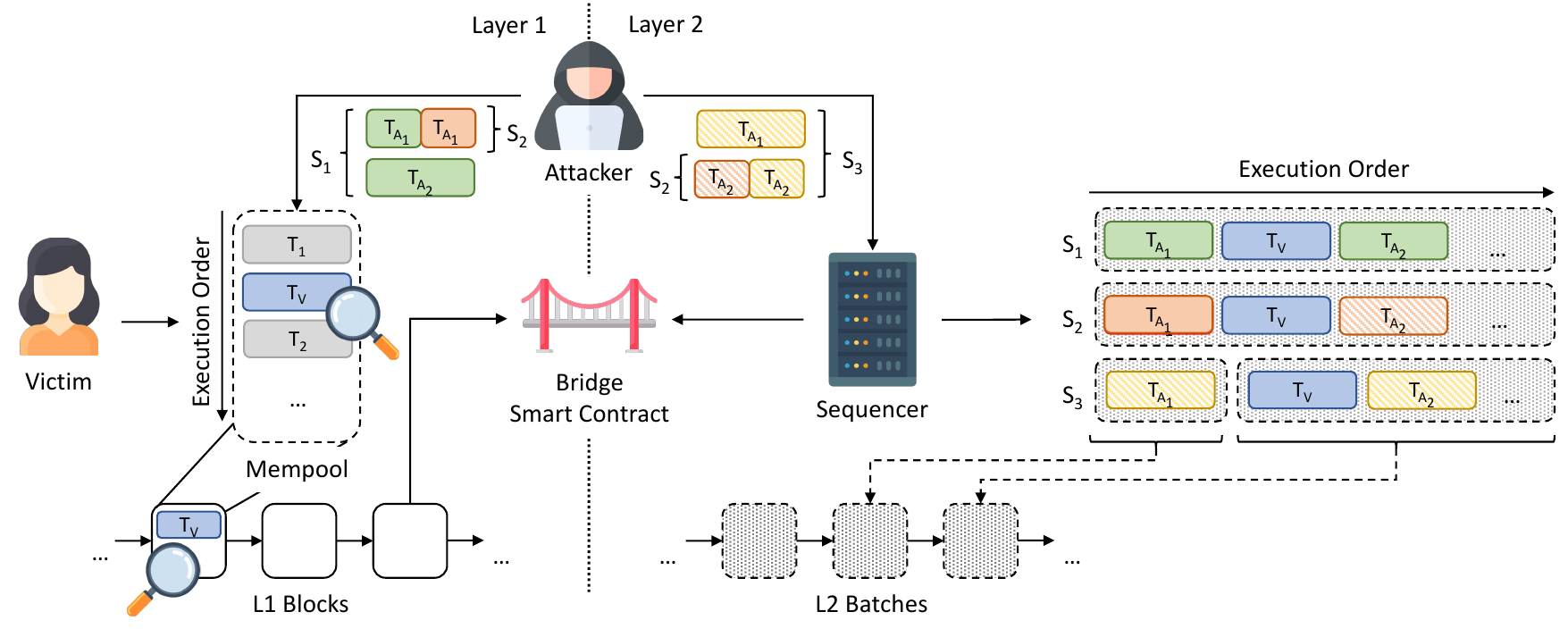}
    \caption{An illustrative example of all three cross-layer sandwich attacks (i.e., $S_1$ highlighted in green, $S_2$ highlighted in orange, and $S_3$ highlighted in yellow) presented in this work. L1 transactions are solid and L2 transactions are striped.}
    \label{fig:attack_startegies}
\end{figure*}

In this section, we propose three novel cross-layer sandwich attacks that leverage the fact that users can send L2 transactions via L1.
We execute these attacks against an Ethereum testnet---a network that runs the Ethereum protocol, but that carries no value and is only used for experiments and tests.
\emph{Even in the testnet, we only target our own transactions.}
We also simulate our attacks on the Ethereum mainnet---the full value Ethereum network---these simulations show how much MEV we could have extracted if we had been running this attack in real time.





\subsection{Attacker Model}

Our model assumes normal blockchain user capabilities (e.g., observing the mempool, sending out transactions, etc.).
An attacker can pursue different strategies to conduct cross-layer sandwich attacks, depending on whether they have access to the mempool on L1 and whether they operate accounts on L1 and L2. 
A user can select transaction fees arbitrarily, but transaction ordering is at the sequencer's discretion on rollups, with different ordering mechanisms discussed in \Cref{tab:rollup_comparison}.
We assume that the sequencer is not colluding with any other users, and that more generally, users can only view transactions in the public mempool.

\subsection{Attack Strategies}
We present the following three cross-layer sandwiching strategies:
    
    \textbf{Strategy $S_1$: Classical Sandwiching.} In this strategy, the attacker performs a classical sandwich attack on L1.  
    Users can send L2 transactions via L1, which as a result are publicly visible in the mempool before finalization.
    The attacker exploits the fact that L1 is vulnerable to frontrunning
    by monitoring the mempool on L1 and searching for transactions that contain L2 transactions that perform token swaps (i.e., $T_{V}$). Afterwards, the attacker performs classical sandwiching by bribing the block builder to include transaction $T_{A_{1}}$ before the observed transaction $T_{V}$, and transaction $T_{A_{2}}$ after the observed transaction $T_{V}$. Despite all transactions being submitted via L1, the actual effect and the resulting price manipulation only occurs on L2 (see green boxes in \figureautorefname{} \ref{fig:attack_startegies}).
    
    \textbf{Strategy $S_2$: Hybrid Sandwiching.} In this strategy, the attacker also monitors the mempool on L1 for L2 transactions that perform token swaps (i.e., $T_{V}$). However, the attacker only performs transaction $T_{A_{1}}$ on L1 and sends out transaction $T_{A_{2}}$ directly to the sequencer on L2. The attacker exploits the fact that sequencers place L2 transactions originating from L1 at the top of L2 batches. This not only provides the attacker with a deterministic backrunning mechanism, but also helps the attacker reducing its costs, as transaction fees are lower on L2 (see orange boxes in \figureautorefname{} \ref{fig:attack_startegies}).
    
    \textbf{Strategy $S_3$: Speculative Sandwiching.} In this strategy, the attacker does not monitor the mempool on L1, but does monitor the transactions included in the latest blocks on L1 and searches for transactions that contain L2 transactions that perform token swaps (i.e., $T_{V}$). The attacker exploits the fact that L2 transactions, which are emitted via L1 transactions, are not immediately included in L2 batches. Instead, the sequencer waits for a timeout period to elapse before extracting the transactions from L1 and including them in an L2 batch. This delay allows the attacker to send the transaction $T_{A_{1}}$ on L2, wait for $T_{V}$ to be included, then send transaction $T_{A_{2}}$ on L2 as well. Since both attacker transactions are emitted via L2, this enables the attacker to further save on transaction fees as compared to strategy $S_1$ (see yellow boxes in \figureautorefname{} \ref{fig:attack_startegies}).

\subsection{Mainnet Simulation}

We simulate our three attack strategies using past mainnet data to analyze their feasibility and measure their potential impact.

\subsubsection{Victim Inference}
\label{sec:victim_inference}


To simulate the impact of our three attacks we first need to find real potential victim transactions. A potential victim is defined as an L2 transaction that is sent via L1 and which performs a trade on a DEX deployed on L2. 
To identify such transactions, we start by scanning past mainnet transactions for L1 transactions which encapsulate L2 transactions. 
By analyzing the bridges deployed by Arbitrum, Optimism, and zkSync, we identified several events that are triggered whenever an L2 transaction is sent via L1. For Arbitrum, an event called \emph{``InboxMessageDelivered''} is emitted, while for Optimism an event called \emph{``TransactionEnqueued''} or \emph{``TransactionDeposited''} is emitted, and for zkSync an event called \emph{``NewPriorityRequest''} is emitted (see Appendix \ref{sec:events}). 
We leverage these events to collect L1 transactions that encapsulate L2 transactions. 

Next, we need to identify L2 counterpart to the L2 transaction that was sent via L1.
This is so that we can determine whether the L2 transaction performs a trade on a DEX on L2. 
For zkSync, this is trivial as the L1 transaction hash is identical to the L2 transaction hash on zkSync. In the case of Arbitrum and Optimism, this is more complicated as the corresponding L2 transaction has a different transaction hash than the L1 transaction.
However, L2 transactions on Arbitrum and Optimism emit specific events when they originate from L1. 
For Arbitrum, we search for L2 transactions which emitted an event called \emph{``RedeemScheduled''} and for Optimism we search for L2 transactions which emitted an event called \emph{``RelayedMessage''}.
As a final step for Arbitrum and Optimism, we need to link the L2 transactions to their corresponding L1 transactions. 
For Arbitrum we use the \emph{message number}, which is a unique sequential identifier generated via Arbitrum's bridge smart contract on L1 and which is part of the event data emitted by the L2 transaction. 
For Optimism, we leverage the \emph{message hash} which is computed as the Keccak hash of the L2 transaction content that was sent via L1.

Once linked, we scan the L2 transactions for DEX swaps. 
A swap typically results in two ERC-20 \emph{``Transfer''} events being emitted. 
Hence, we scan the events emitted by the identified L2 transactions and mark these as potential victims if we find two \emph{``Transfer''} events emitted by the same token smart contract, where the sender and receiver of the tokens are swapped. 
We found a total of \num{1916524} L2 transactions emitted via L1, out of which \num{1459570} were observed on Arbitrum, \num{424212} on Optimism, and \num{32742} on zkSync.
From the observed 1.9M L2 transactions, only \num{170674} ($\approx$ 9\%) can be considered potential victims based on our methodology to detect token swaps, where \num{87921} are from Arbitrum and \num{82753} from Optimism. We could not find potential victims on zkSync. This is most likely due to the fact that there was no cross-layer DEX currently deployed on zkSync at the time of our measurement. 
\tableautorefname{} \ref{tab:victims} in Appendix \ref{sec:victims} provides a detailed overview of the identified DEXes and their liquidity pools on Arbitrum and Optimism. 
We observe that essentially all potential victim transactions originate from users using the Hop Protocol \cite{hop_protocol}. The Hop protocol allows users to send tokens from one rollup or chain to another by employing their own market makers, hence making it a perfect victim for our cross-layer sandwich attacks.

\begin{figure}
    \centering
    \includegraphics[width=\columnwidth]{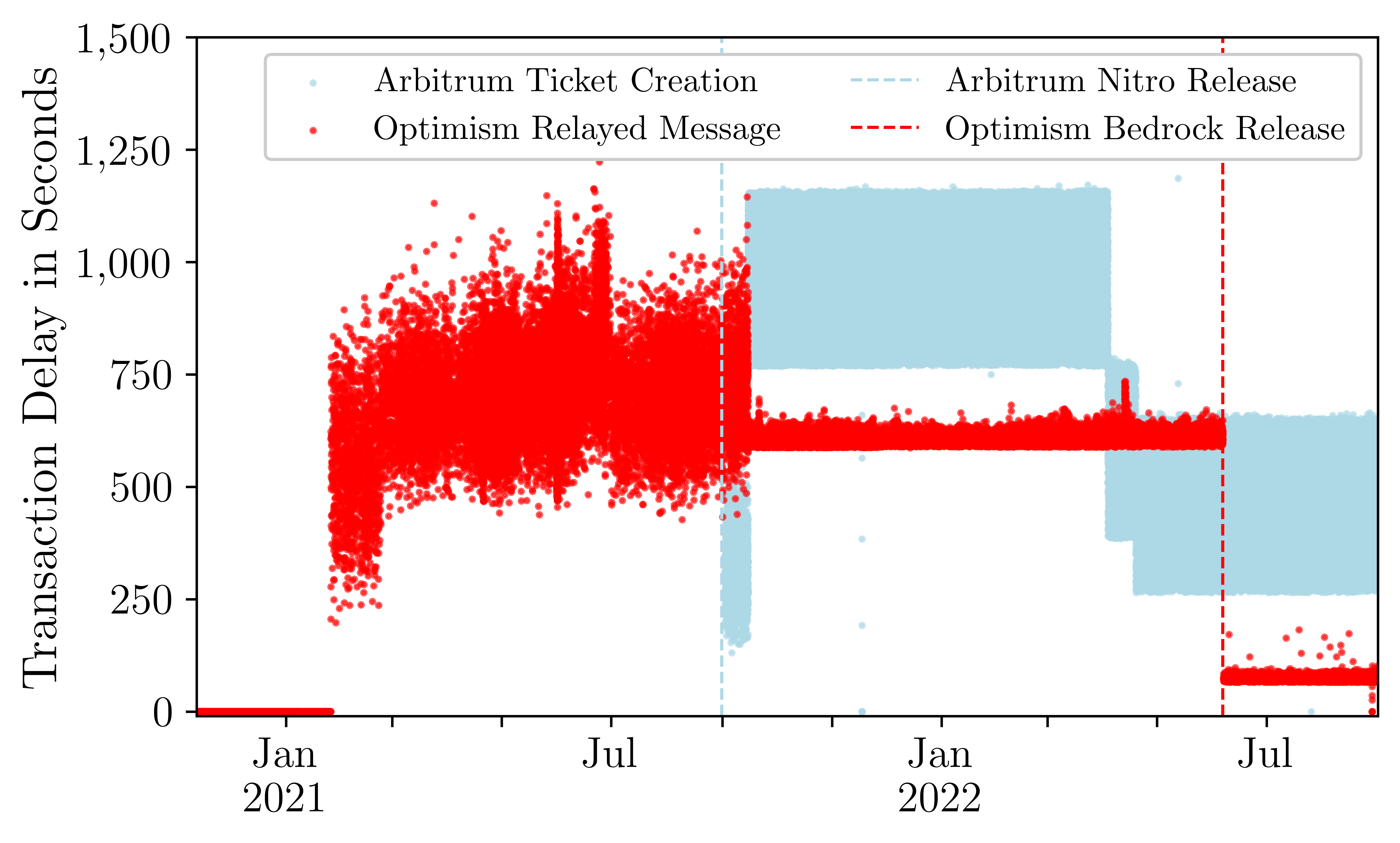}
    \caption{Delay in seconds between L1 transactions and L2 transactions on Arbitrum and Optimism.}
    \label{fig:message_delay}
\end{figure}

\subsubsection{Transaction Inclusion Delays} As mentioned earlier, L2 transactions that are sent via L1 are not directly included in L2 batches by the sequencer. The sequencer typically does this only after a certain time has passed. This is to obtain some confirmation on the finality of the observed L1 transactions, as their inclusion might still change early on due to uncle blocks or chain reorganizations (i.e., reorgs). 
We compared the timestamps of L1 transactions and their corresponding L2 transactions of the identified potential victims to understand the delay that is imposed by the sequencers. 
This allows us to determine whether attack strategy $S_3$ is feasible in practice. 
If the delay is too short or zero, then attackers do not have enough time to react to emitted L2 transactions that they observed via L1 blocks.
We computed the delay by subtracting the L1 transaction timestamp from the L2 transaction timestamp of the previously identified victim transactions to see how long it takes for an L1 transaction to be included on L2. \figureautorefname{} \ref{fig:message_delay} depicts our results.

For Arbitrum, we find that the smallest delay (i.e., the fastest inclusion of an L1 transaction) was 0 seconds, whereas the mean is around 798 seconds ($\approx$ 13 minutes) and the median is around 868 seconds ($\approx$ 14 minutes).
The maximum that we observed (i.e., the slowest inclusion of an L1 transaction) is \num{3694} seconds ($\approx$ 1 hour). 
In \figureautorefname{} \ref{fig:message_delay},  we see that the delay usually stays within some boundaries and that these boundaries have moved over time towards shorter delays. However, we can also observe that a delay of zero seconds is typically an outlier and that almost always the delay is at least 120 seconds (i.e.,  minutes). This means that, on Arbitrum, attackers typically have a window of at least 2 minutes to send their frontrunning transaction $T_{A_1}$ after observing the L2 transaction on L1. Theoretically, attackers can have even more time since the delays that we presented here  are related to the ticket creation transactions and not the ticket redeem transactions (i.e., which actually execute the victim's L2 transaction). However, it is up to the user/protocol to send the redeem transactions, after the ticket creation transactions have been included. This means that the smallest possible delay is near zero seconds after the ticket creation. Therefore, the ticket creation transaction delay is a good reference point to measure the delay that an attacker can leverage.

For Optimism, we can compare transaction delays before and after the Bedrock release (i.e., version 2.0). For both releases (i.e., pre- and post-Bedrock), we find that the minimum is zero seconds (i.e., instant inclusion). 
However, in \figureautorefname{} \ref{fig:message_delay}, we observe that delays of zero seconds occurred only in the earlier days of Optimism's existence and that it changed around February 2022. 
For the post-Bedrock release the zero second delays seem to be outliers. 
Pre-Bedrock, the mean and median delays were around 583 seconds ($\approx$ 9 minutes) and 609 seconds ($\approx$ 10 minutes) respectively. 
While post-Bedrock, the mean and median delays are around 83 seconds and 74 seconds ($\approx$ 1 minute), respectively. 
The maximum (i.e., largest delay) that we observed pre- and post-Bedrock was \num{1266} seconds and \num{22296}, respectively. 
Similar to Arbitrum, the delays seem to stay within certain bounds.
In contrast to Arbitrum, there is more variation on Optimism, though it decreased over time.


\begin{table}[b]
    \centering
    \begin{tabular}{l | R{0.8cm} R{0.8cm} R{0.8cm} | R{0.8cm} R{0.8cm} R{0.8cm}}
        \toprule
        & \multicolumn{3}{c}{\textbf{Arbitrum}} & \multicolumn{3}{c}{\textbf{Optimism}} \\
        \textbf{Capital} & $S_1$ & $S_2$ & $S_3$ & $S_1$ & $S_2$ & $S_3$ \\
        \midrule
        1K &  10 & 17 & 216 & 7 & 29 & 709 \\
        10K & 86 & 126 & 254 & 196 & 302 & 815 \\
        100K & 220 & 237 & 277 & 534 & 643 & 847 \\
        1M & 289 & 295 & 320 & 773 & 845 & 965 \\
        $\infty$ & 314 & 322 & 345 & 825 & 898 & \num{1015} \\
        \bottomrule 
    \end{tabular}
    \caption{Profitable victim transactions per attack strategy and attacker capital.}
    \label{tab:profitable_transactions}
\end{table}

\subsubsection{Estimated Profits.} We calculate potential profits by simulating our three attack strategies using the previously identified victim transactions. 
For each victim transaction, we extract the amount that is being traded. The attacker’s profit hinges on maximizing its purchase amount within the victim’s slippage threshold without exceeding it.
We compute the optimal amount using a ternary search algorithm as suggested by other works \cite{HeimbachW22} and use this amount to simulate and compute the gain of performing a cross-layer sandwich attack against the victim transaction. 
The final profit is calculated for each attack strategy by deducting the individual transaction costs (which depend on the employed attack strategy) from the gained amount. 
Moreover, since sandwich attacks require the a priori purchase of assets in $T_{A_1}$, attackers are required to posses a certain upfront capital.
To gauge the profitability across various capital ranges, we consider budgets of 1K USD, 10K USD, 100K USD, 1M USD, and $\infty$-capital. 

\begin{figure}
    \centering
    \includegraphics[width=\columnwidth]{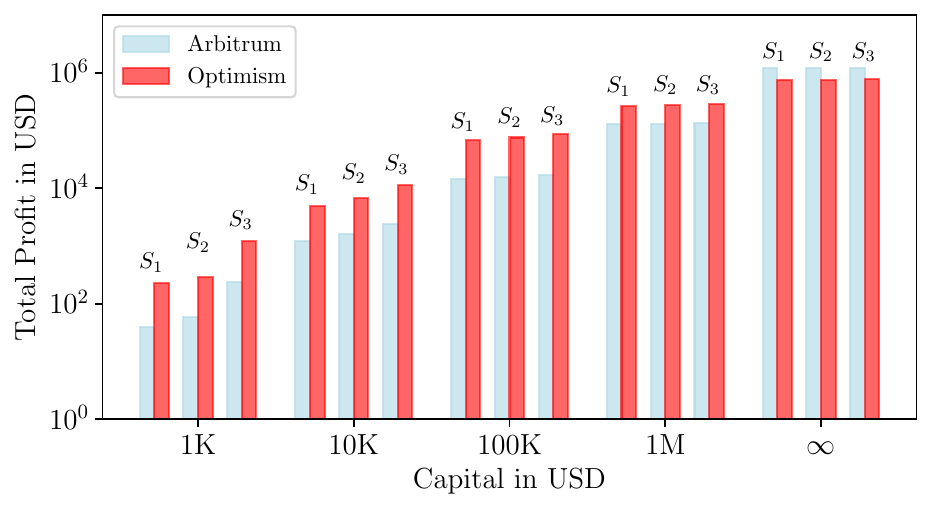}
    \caption{Total profit in USD on Arbitrum and Optimism for each attack strategy across different attacker capitals.}
    \label{fig:cross_layer_sandwich_profits}
\end{figure}

\tableautorefname{} \ref{tab:sandwich_profitability} in Appendix \ref{sec:appendix_d} provides a detailed overview on the profits obtained for each attack strategy across Arbitrum and Optimism using different attacker budgets. 
Interestingly, not all identified potential victim transactions are profitable. 
These transactions are either too small to offer revenue outweighing costs, or have low slippage tolerance settings which do not provide enough room for frontrunning. 
In \tableautorefname{} \ref{tab:profitable_transactions}, we observe that the number of profitable transactions increases together with the attacker's capital, suggesting that several victim transactions require large purchases by attackers in order to be profitable. 
Nonetheless, we can state that the maximum total profit is 1.2M USD for Arbitrum and 767K USD for Optimism. 
\figureautorefname{} \ref{fig:cross_layer_sandwich_profits} further highlights that Optimism is more profitable than Arbitrum for attackers that posses a capital up to 1M USD. 
In addition, we observe that attack strategy $S_3$ is always more profitable than $S_2$ and $S_1$.
However, the difference diminishes as the attacker capital increases. 
Hence, we conclude that either strategies are profitable, but that for attackers with smaller capital (e.g., 1K USD) strategy $S_3$ is more profitable than for attackers with larger capital (e.g., 1M USD).

\subsection{Testnet Validation}

While our mainnet simulation provides insights such as number of potential victims and estimated profits, it does not validate the deployability of our attacks in practice. 
To that end, we leverage live test networks to demonstrate the practicality of our cross-layer sandwich attacks.
Given the similarities between testnets and mainnets, we can validate the feasibility of our attacks in an ethical manner
without impacting real users and requiring substantial financial resources. 
For our analysis, we used the Sepolia testnet  alongside its corresponding L2 counterparts for Arbitrum, Optimism, and zkSync. 
We developed a script to automate the process of deploying our own attacker and victim accounts on each chain including our own Uniswap V2-inspired DEXes on Arbitrum, Optimism, and zkSync. 
\emph{The only victim was our own account even on the testnet.}
The script automatically performs the three attack strategies. First, a victim transaction $T_V$ is sent via the official L1 bridge of the rollup that is under test. The victim transaction simply performs a swap on the Uniswap-alike DEX that the script previously deployed on L2. Afterwards, the script simulates the attacker transactions $T_{A_1}$ and $T_{A_2}$ according to the selected strategy. The script either scans the mempool or the latest blocks on L1 and searches for our victim transaction $T_V$ as detailed in Section \ref{sec:victim_inference} and crafts the corresponding transactions $T_{A_1}$ and $T_{A_2}$, which essentially buy and sell the same asset as $T_V$ on the Uniswap-alike DEX on L2. 
See \tableautorefname{} \ref{tab:testnet_sandwiching} in \Cref{sec:appendix_e} for a detailed list of our successfully deployed attack transactions across Arbitrum, Optimism, and zkSync.

\section{Discussion}

In this section, we discuss the generalization as well as limitations of our cross-layer sandwich attacks and potential countermeasures.

\subsection{Generalization and Limitations}

We performed our measurements on Arbitrum, Optimism, and zkSync, as these were the most popular rollups by market share at the time of writing \cite{l2beat}. However, since then other rollups such as Base \cite{base} and Blast \cite{blast} have gained tremendously on popularity. As these rollups are based on Optimism's OP Stack or Superchain Ecosystem \cite{superchain}, our scripts to detect MEV on Optimism can easily be repurposed to measure MEV extraction on any Optimism fork. 
While we use sandwich attacks as a case study, our proposed strategies $S_1$, $S_2$, and $S_3$ can be further generalized to cross-layer frontrunning attacks. We posit that our work applies any time there is a victim transaction making L1-to-L2 transactions on a blockchain with a public mempool (strategies $S_1$ and $S_2$), or when a rollup imposes a delay in including L1 transactions (strategy $S_3$). Sandwich attacks exploit both frontrunning and backrunning, so it is possible that we could see our strategies applied to other techniques besides sandwiching that would also yield profit when transactions are frontrun (e.g., generalized frontrunners \cite{exploitZhang2023,qin2023blockchain}).
All of our proposed cross-layer sandwich attacks exploit the fact that users submit L2 transactions via L1. 
While this seems counter-intuitive, there exist legitimate use cases which require such behavior. 
For example, they are used in cross-chain swaps as provided via the Hop Protocol \cite{hop_protocol} or the Across Protocol \cite{across_protocol}. 
The reason why we were only able to find victim transactions related to Hop, was due to the fact that Across leverages their own bridges and does not rely on the native bridges deployed by individual rollup technologies. 
However, Across's bridge also hinges on emitting events to trigger cross-chain swaps, hence, our victim inference could be adjusted for Across as well. 
Strategies $S_1$ and $S_2$ depend on the fact that attackers can observe victim transactions via the mempool.
Hence, any L1 blockchain that has a public mempool is susceptible to these two strategies. 
Finally, we did not find any victim transactions on zkSync due to the fact that during our data collection period no cross-chain exchange such as Hop was operating.

\subsection{Countermeasures}

Private pools \cite{flashbots}, including encrypted mempools \cite{kavousi2023blindperm}, provide an effective countermeasure against strategies $S_1$ and $S_2$ since attackers are not able to observe victim transactions via the mempool anymore and thus cannot frontrun them. 
However, strategy $S_3$ still remains feasible as it does not rely on victim transactions being visible before a block is finalized on L1. 
Strategy $S_3$ exploits the fact that there is a delay between L1 and L2 transactions. 
Rollups could try to reduce the delay almost to zero, such that attackers cannot react fast enough. However, this would make rollups vulnerable to time bandit attacks or uncle bandit attacks, which are already viable issues on Ethereum \cite{unmasking_bandit_attacks}. Another solution could be randomized ordering of transactions. 
However, this would break the first-come, first-served ordering policy of many rollups and would not entirely solve the issue of sandwich attacks as shown by previous works \cite{tumas2023ripple}.
Finally, a number of works aim to counter sandwiching on AMMs by proposing new frontrunning resistant AMM protocols \cite{zhou2021a2mm,heimbach2022eliminating,wadhwa2023data}. However, these protocols were exclusively designed to protect AMMs against frontrunning and not any other protocols. Moreover, it is not clear whether these protocols are still applicable in the context of cross-layer sandwiching.

\section{Related Work}

\citet{gudgeon2020sok} presented the first SoK on Layer-2, however, their discussion was before the advent of rollups.
The first academic work on MEV was done by \citet{flashboys}. \citet{ferreira2021frontrunner} and \citet{Qin_Zhou_Gervais_2022} were the first to measure the historical prevalence of MEV on the blockchain. \citet{Piet_Fairoze_Weaver_2022} analyzed private pool usage by measuring mempool transactions from several vantage points, but did not study MEV.
\citet{flashbotinthepan} used related methods to measure the prevalence of MEV in private pools especially on Flashbots~\cite{flashbots}.
While these were important works for identifying the widespread phenomenon of MEV, their  measurements were before the advent of rollups and, hence, only focused on Ethereum.


The first work to study MEV on rollups was by \citet{flashbabies}, which measured MEV on Optimism and Polygon.
These measurements, however, persisted for only a couple months and did not consider Arbitrum or any ZK rollups which had not been released yet.
Also at the intersection of MEV and rollups, \citet{bagourd2023quantifying} scanned Polygon, Optimism, and Arbitrum for MEV. However, they did not scan for sandwiches, or analyze the negative aspects of MEV, nor did they compare their results to Ethereum. Besides providing results for a longer measurement period, our methodology has found more instances of MEV for the same time period than \citet{bagourd2023quantifying} reported. Recently, \citet{oz2024playing} have studied MEV extraction on Algorand, which follows a first-come, first-served policy similar to rollups.

In the context of cross-chain MEV, \citet{obadia2021unity} contributed a formalization of cross-domain MEV, but their definitions do not directly yield any method for quickly finding new instances of cross-chain MEV.
In a similar vein, \citet{zhang2023front} defined cross-shard frontrunning and contributed a model for defending against it.
These, however, are orthogonal to the frontrunning issue in rollups.

\citet{McLaughlinKV23} present a different technique to detect arbitrages as compared to existing works that is more application agnostic.
\citet{li2023demystifying} characterize several novel forms of MEV and measure their appearance within Flashbots bundles.
Likewise, \citet{babel2023lanturn} leverage machine learning to learn new strategies to extract MEV---their model is capable of detecting two new strategies. However, neither of the aforementioned works considers MEV in the context of rollups.
Moreover, besides comparing MEV extraction between Ethereum and rollups, we also present three novel strategies that enable users to perform sandwich attacks on rollups.


\section{Conclusion}

MEV is omnipresent, even on rollups. We analyzed the extraction of MEV across Ethereum, Arbitrum, Optimism, and zkSync for a period of nearly three years. We found a significant amount of arbitrages and liquidations occurring on rollups. 
Compared to Ethereum, and despite similar volume and lower costs, MEV on rollups yield smaller profits. 
The absence of proposer-builder separation in rollups reflects in the frequency of failed transactions by competitors. 
Moreover, MEV opportunities take longer to materialize on rollups as compared to Ethereum. 
Nonetheless, we found instances where extraction and opportunity coincided on rollups, hinting at potential for shorter extraction delays.
While no traditional sandwiching was uncovered on rollups, we detected three potential cross-layer sandwich attacks. 
These exploit L2 transactions routed via L1 to execute sandwiching on rollups, sidestepping the need for a public mempool on rollups as it is required for traditional sandwiching. 
Assessing these attacks, we estimated potential profits nearing 2 million USD. 
This underscores the necessity for further research to comprehend MEV's impact on cross-chain communication.

\begin{acks}
This work was supported by the Zurich Information Security \& Privacy Center (ZISC).
This work is also supported by the European Union’s Horizon 2020 research and innovation programme under grant agreement No 952226, project BIG (Enhancing the research and innovation potential of Tecnico through Blockchain technologies and design Innovation for social Good) as well as by national funds through FCT, Fundação para a
Ciência e a Tecnologia, under projects UIDB/50021/2020
(DOI:10.54499/UIDB/50021/2020) and UIDP/50021/2020
(DOI:10.54499/UIDP/50021/2020).
\end{acks}

\bibliographystyle{ACM-Reference-Format}
\balance
\bibliography{references}


\begin{thebibliography}{75}


\ifx \showCODEN    \undefined \def \showCODEN     #1{\unskip}     \fi
\ifx \showDOI      \undefined \def \showDOI       #1{#1}\fi
\ifx \showISBNx    \undefined \def \showISBNx     #1{\unskip}     \fi
\ifx \showISBNxiii \undefined \def \showISBNxiii  #1{\unskip}     \fi
\ifx \showISSN     \undefined \def \showISSN      #1{\unskip}     \fi
\ifx \showLCCN     \undefined \def \showLCCN      #1{\unskip}     \fi
\ifx \shownote     \undefined \def \shownote      #1{#1}          \fi
\ifx \showarticletitle \undefined \def \showarticletitle #1{#1}   \fi
\ifx \showURL      \undefined \def \showURL       {\relax}        \fi
\providecommand\bibfield[2]{#2}
\providecommand\bibinfo[2]{#2}
\providecommand\natexlab[1]{#1}
\providecommand\showeprint[2][]{arXiv:#2}

\bibitem[Aave(2024a)]%
        {aave}
\bibfield{author}{\bibinfo{person}{Aave}.} \bibinfo{year}{2024}\natexlab{a}.
\newblock \bibinfo{title}{Aave - Open Source Liquidity Protocol}.
\newblock
\newblock
\urldef\tempurl%
\url{https://aave.com/}
\showURL{%
\tempurl}
\newblock
\shownote{Online; accessed 20 April 2024}.


\bibitem[Aave(2024b)]%
        {aave_flash_loan}
\bibfield{author}{\bibinfo{person}{Aave}.} \bibinfo{year}{2024}\natexlab{b}.
\newblock \bibinfo{title}{Flash Loans - Developers}.
\newblock
\newblock
\urldef\tempurl%
\url{https://docs.aave.com/developers/guides/flash-loans}
\showURL{%
\tempurl}
\newblock
\shownote{Online; accessed 20 April 2024}.


\bibitem[Aave(2024c)]%
        {health_factor_aave}
\bibfield{author}{\bibinfo{person}{Aave}.} \bibinfo{year}{2024}\natexlab{c}.
\newblock \bibinfo{title}{LendingPool - Developers}.
\newblock
\newblock
\urldef\tempurl%
\url{https://docs.aave.com/developers/v/2.0/the-core-protocol/lendingpool#getuseracountdata}
\showURL{%
\tempurl}
\newblock
\shownote{Online; accessed 20 April 2024}.


\bibitem[Aave(2024d)]%
        {compound_liquidations}
\bibfield{author}{\bibinfo{person}{Aave}.} \bibinfo{year}{2024}\natexlab{d}.
\newblock \bibinfo{title}{Liquidations - Developers}.
\newblock
\newblock
\urldef\tempurl%
\url{https://docs.aave.com/developers/guides/liquidations}
\showURL{%
\tempurl}
\newblock
\shownote{Online; accessed 20 April 2024}.


\bibitem[Arbiscan(2024)]%
        {arbiscan_api}
\bibfield{author}{\bibinfo{person}{Arbiscan}.} \bibinfo{year}{2024}\natexlab{}.
\newblock \bibinfo{title}{Accounts | Arbiscan}.
\newblock
\newblock
\urldef\tempurl%
\url{https://docs.arbiscan.io/api-endpoints/accounts}
\showURL{%
\tempurl}
\newblock
\shownote{Online; accessed 20 April 2024}.


\bibitem[Arbitrum(2024)]%
        {arbitrum}
\bibfield{author}{\bibinfo{person}{Arbitrum}.} \bibinfo{year}{2024}\natexlab{}.
\newblock \bibinfo{title}{Arbitrum — The Future of Ethereum}.
\newblock
\newblock
\urldef\tempurl%
\url{https://arbitrum.io/}
\showURL{%
\tempurl}
\newblock
\shownote{Online; accessed 20 April 2024}.


\bibitem[Babel et~al\mbox{.}(2023)]%
        {babel2023lanturn}
\bibfield{author}{\bibinfo{person}{Kushal Babel}, \bibinfo{person}{Mojan
  Javaheripi}, \bibinfo{person}{Yan Ji}, \bibinfo{person}{Mahimna Kelkar},
  \bibinfo{person}{Farinaz Koushanfar}, {and} \bibinfo{person}{Ari Juels}.}
  \bibinfo{year}{2023}\natexlab{}.
\newblock \showarticletitle{Lanturn: Measuring economic security of smart
  contracts through adaptive learning}. In
  \bibinfo{booktitle}{\emph{Proceedings of the 2023 ACM SIGSAC Conference on
  Computer and Communications Security}}. \bibinfo{pages}{1212--1226}.
\newblock


\bibitem[Bagourd and Francois(2023)]%
        {bagourd2023quantifying}
\bibfield{author}{\bibinfo{person}{Arthur Bagourd} {and}
  \bibinfo{person}{Luca~Georges Francois}.} \bibinfo{year}{2023}\natexlab{}.
\newblock \showarticletitle{Quantifying {MEV} On Layer 2 Networks}.
\newblock \bibinfo{journal}{\emph{CoRR}} (\bibinfo{year}{2023}).
\newblock
\urldef\tempurl%
\url{https://doi.org/10.48550/arXiv.2309.00629}
\showURL{%
\tempurl}


\bibitem[Balancer(2024a)]%
        {balancer}
\bibfield{author}{\bibinfo{person}{Balancer}.}
  \bibinfo{year}{2024}\natexlab{a}.
\newblock \bibinfo{title}{Balancer DeFi Liquidity Protocol}.
\newblock
\newblock
\urldef\tempurl%
\url{https://balancer.fi/}
\showURL{%
\tempurl}
\newblock
\shownote{Online; accessed 20 April 2024}.


\bibitem[Balancer(2024b)]%
        {balancer_flash_loan}
\bibfield{author}{\bibinfo{person}{Balancer}.}
  \bibinfo{year}{2024}\natexlab{b}.
\newblock \bibinfo{title}{Flash Loans | Balancer}.
\newblock
\newblock
\urldef\tempurl%
\url{https://docs.balancer.fi/reference/contracts/flash-loans.html}
\showURL{%
\tempurl}
\newblock
\shownote{Online; accessed 20 April 2024}.


\bibitem[Blast(2024)]%
        {blast}
\bibfield{author}{\bibinfo{person}{Blast}.} \bibinfo{year}{2024}\natexlab{}.
\newblock \bibinfo{title}{About Blast - Blast Developer Documentation}.
\newblock
\newblock
\urldef\tempurl%
\url{https://docs.blast.io/about-blast}
\showURL{%
\tempurl}
\newblock
\shownote{Online; accessed 5 September 2024}.


\bibitem[Chain(2024)]%
        {ronin}
\bibfield{author}{\bibinfo{person}{Ronin Chain}.}
  \bibinfo{year}{2024}\natexlab{}.
\newblock \bibinfo{title}{Ronin Bridge}.
\newblock
\newblock
\urldef\tempurl%
\url{https://docs.roninchain.com/apps/ronin-bridge}
\showURL{%
\tempurl}
\newblock
\shownote{Online; accessed 20 April 2024}.


\bibitem[Chainlink(2024)]%
        {chainlink_price_oracle}
\bibfield{author}{\bibinfo{person}{Chainlink}.}
  \bibinfo{year}{2024}\natexlab{}.
\newblock \bibinfo{title}{Chainlink Data Feeds | Chainlink Documentation}.
\newblock
\newblock
\urldef\tempurl%
\url{https://docs.chain.link/data-feeds#price-feeds}
\showURL{%
\tempurl}
\newblock
\shownote{Online; accessed 20 April 2024}.


\bibitem[Coinbase(2024)]%
        {base}
\bibfield{author}{\bibinfo{person}{Coinbase}.} \bibinfo{year}{2024}\natexlab{}.
\newblock \bibinfo{title}{Base}.
\newblock
\newblock
\urldef\tempurl%
\url{https://www.base.org/}
\showURL{%
\tempurl}
\newblock
\shownote{Online; accessed 5 September 2024}.


\bibitem[CoinGecko(2024)]%
        {coingecko_api}
\bibfield{author}{\bibinfo{person}{CoinGecko}.}
  \bibinfo{year}{2024}\natexlab{}.
\newblock \bibinfo{title}{Crypto API Documentation | CoinGecko}.
\newblock
\newblock
\urldef\tempurl%
\url{https://www.coingecko.com/api/documentation}
\showURL{%
\tempurl}
\newblock
\shownote{Online; accessed 20 April 2024}.


\bibitem[CompaniesMarketCap.com(2024a)]%
        {companiesmarketcap}
\bibfield{author}{\bibinfo{person}{CompaniesMarketCap.com}.}
  \bibinfo{year}{2024}\natexlab{a}.
\newblock \bibinfo{title}{Companies ranked by Market Cap -
  CompaniesMarketCap.com}.
\newblock
\newblock
\urldef\tempurl%
\url{https://companiesmarketcap.com/}
\showURL{%
\tempurl}
\newblock
\shownote{Online; accessed 20 April 2024}.


\bibitem[CompaniesMarketCap.com(2024b)]%
        {defillama_chains}
\bibfield{author}{\bibinfo{person}{CompaniesMarketCap.com}.}
  \bibinfo{year}{2024}\natexlab{b}.
\newblock \bibinfo{title}{Total Value Locked All Chains - DeFiLama}.
\newblock
\newblock
\urldef\tempurl%
\url{https://defillama.com/chains}
\showURL{%
\tempurl}
\newblock
\shownote{Online; accessed 20 April 2024}.


\bibitem[Compound(2024a)]%
        {compound}
\bibfield{author}{\bibinfo{person}{Compound}.}
  \bibinfo{year}{2024}\natexlab{a}.
\newblock \bibinfo{title}{Compound}.
\newblock
\newblock
\urldef\tempurl%
\url{https://compound.finance/}
\showURL{%
\tempurl}
\newblock
\shownote{Online; accessed 20 April 2024}.


\bibitem[Compound(2024b)]%
        {shortfall_compound}
\bibfield{author}{\bibinfo{person}{Compound}.}
  \bibinfo{year}{2024}\natexlab{b}.
\newblock \bibinfo{title}{Compound V2 Docs | Comptroller}.
\newblock
\newblock
\urldef\tempurl%
\url{https://docs.compound.finance/v2/comptroller/}
\showURL{%
\tempurl}
\newblock
\shownote{Online; accessed 20 April 2024}.


\bibitem[Curve(2024)]%
        {curve}
\bibfield{author}{\bibinfo{person}{Curve}.} \bibinfo{year}{2024}\natexlab{}.
\newblock \bibinfo{title}{Curve: Swap}.
\newblock
\newblock
\urldef\tempurl%
\url{https://curve.fi}
\showURL{%
\tempurl}
\newblock
\shownote{Online; accessed 20 April 2024}.


\bibitem[Daian et~al\mbox{.}(2020)]%
        {flashboys}
\bibfield{author}{\bibinfo{person}{Philip Daian}, \bibinfo{person}{Steven
  Goldfeder}, \bibinfo{person}{Tyler Kell}, \bibinfo{person}{Yunqi Li},
  \bibinfo{person}{Xueyuan Zhao}, \bibinfo{person}{Iddo Bentov},
  \bibinfo{person}{Lorenz Breidenbach}, {and} \bibinfo{person}{Ari Juels}.}
  \bibinfo{year}{2020}\natexlab{}.
\newblock \showarticletitle{Flash Boys 2.0: Frontrunning in Decentralized
  Exchanges, Miner Extractable Value, and Consensus Instability}. In
  \bibinfo{booktitle}{\emph{2020 IEEE Symposium on Security and Privacy (SP)}}.
  \bibinfo{pages}{910–927}.
\newblock
\showISSN{2375-1207}


\bibitem[Dang et~al\mbox{.}(2019)]%
        {dang2019towards}
\bibfield{author}{\bibinfo{person}{Hung Dang}, \bibinfo{person}{Tien Tuan~Anh
  Dinh}, \bibinfo{person}{Dumitrel Loghin}, \bibinfo{person}{Ee-Chien Chang},
  \bibinfo{person}{Qian Lin}, {and} \bibinfo{person}{Beng~Chin Ooi}.}
  \bibinfo{year}{2019}\natexlab{}.
\newblock \showarticletitle{Towards scaling blockchain systems via sharding}.
  In \bibinfo{booktitle}{\emph{Proceedings of the 2019 international conference
  on management of data}}. \bibinfo{pages}{123--140}.
\newblock


\bibitem[DappRadar(2024)]%
        {popular_dexes}
\bibfield{author}{\bibinfo{person}{DappRadar}.}
  \bibinfo{year}{2024}\natexlab{}.
\newblock \bibinfo{title}{Top Ethereum DeFi TVL}.
\newblock
\newblock
\urldef\tempurl%
\url{https://dappradar.com/rankings/defi/chain/ethereum?category=defi_dex}
\showURL{%
\tempurl}
\newblock
\shownote{Online; accessed 20 April 2024}.


\bibitem[DeFiLama(2024)]%
        {defillama}
\bibfield{author}{\bibinfo{person}{DeFiLama}.} \bibinfo{year}{2024}\natexlab{}.
\newblock \bibinfo{title}{DeFiLama - DeFi Dashboard}.
\newblock
\newblock
\urldef\tempurl%
\url{https://defillama.com/}
\showURL{%
\tempurl}
\newblock
\shownote{Online; accessed 20 April 2024}.


\bibitem[DeFiLlama(2024)]%
        {popular_lending_platforms}
\bibfield{author}{\bibinfo{person}{DeFiLlama}.}
  \bibinfo{year}{2024}\natexlab{}.
\newblock \bibinfo{title}{Lending TVL Rankings}.
\newblock
\newblock
\urldef\tempurl%
\url{https://defillama.com/protocols/Lending/Ethereum}
\showURL{%
\tempurl}
\newblock
\shownote{Online; accessed 20 April 2024}.


\bibitem[Etherscan(2024a)]%
        {etherscan_api}
\bibfield{author}{\bibinfo{person}{Etherscan}.}
  \bibinfo{year}{2024}\natexlab{a}.
\newblock \bibinfo{title}{Accounts | Etherscan}.
\newblock
\newblock
\urldef\tempurl%
\url{https://docs.etherscan.io/api-endpoints/accounts}
\showURL{%
\tempurl}
\newblock
\shownote{Online; accessed 20 April 2024}.


\bibitem[Etherscan(2024b)]%
        {optimistic_etherscan_api}
\bibfield{author}{\bibinfo{person}{Optimistic Etherscan}.}
  \bibinfo{year}{2024}\natexlab{b}.
\newblock \bibinfo{title}{Accounts | Optimism Etherscan | Optimism}.
\newblock
\newblock
\urldef\tempurl%
\url{https://docs.optimism.etherscan.io/api-endpoints/accounts}
\showURL{%
\tempurl}
\newblock
\shownote{Online; accessed 20 April 2024}.


\bibitem[Flashbots(2024a)]%
        {flashbots}
\bibfield{author}{\bibinfo{person}{Flashbots}.}
  \bibinfo{year}{2024}\natexlab{a}.
\newblock \bibinfo{title}{Flashbots}.
\newblock
\newblock
\urldef\tempurl%
\url{https://www.flashbots.net}
\showURL{%
\tempurl}
\newblock
\shownote{Online; accessed 20 April 2024}.


\bibitem[Flashbots(2024b)]%
        {flashbots_api}
\bibfield{author}{\bibinfo{person}{Flashbots}.}
  \bibinfo{year}{2024}\natexlab{b}.
\newblock \bibinfo{title}{Flashbots Blocks API}.
\newblock
\newblock
\urldef\tempurl%
\url{https://blocks.flashbots.net}
\showURL{%
\tempurl}
\newblock
\shownote{Online; accessed 20 April 2024}.


\bibitem[Gudgeon et~al\mbox{.}(2020)]%
        {gudgeon2020sok}
\bibfield{author}{\bibinfo{person}{Lewis Gudgeon}, \bibinfo{person}{Pedro
  Moreno-Sanchez}, \bibinfo{person}{Stefanie Roos}, \bibinfo{person}{Patrick
  McCorry}, {and} \bibinfo{person}{Arthur Gervais}.}
  \bibinfo{year}{2020}\natexlab{}.
\newblock \showarticletitle{Sok: Layer-two blockchain protocols}. In
  \bibinfo{booktitle}{\emph{Financial Cryptography and Data Security: 24th
  International Conference, FC 2020, Kota Kinabalu, Malaysia, February 10--14,
  2020 Revised Selected Papers 24}}. Springer, \bibinfo{pages}{201--226}.
\newblock


\bibitem[Ha et~al\mbox{.}(2021)]%
        {flashbabies}
\bibfield{author}{\bibinfo{person}{Huy Ha}, \bibinfo{person}{Vasiliki Vlachou},
  \bibinfo{person}{Quintus Kilbourn}, {and} \bibinfo{person}{Cesare
  De~Michellis}.} \bibinfo{year}{2021}\natexlab{}.
\newblock \bibinfo{title}{Flashbabies: Mev on l2}.
\newblock
\newblock
\urldef\tempurl%
\url{https://timroughgarden.github.io/fob21/reports/r11.pdf}
\showURL{%
\tempurl}


\bibitem[Halpern(2021)]%
        {unmasking_bandit_attacks}
\bibfield{author}{\bibinfo{person}{Elan Halpern}.}
  \bibinfo{year}{2021}\natexlab{}.
\newblock \bibinfo{title}{Unmasking the Ethereum Uncle Bandit}.
\newblock
\newblock
\urldef\tempurl%
\url{https://medium.com/alchemy-api/unmasking-the-ethereum-uncle-bandit-a2b3eb694019}
\showURL{%
\tempurl}
\newblock
\shownote{Online; accessed 20 April 2024}.


\bibitem[Heimbach et~al\mbox{.}(2023)]%
        {HeimbachKTW23}
\bibfield{author}{\bibinfo{person}{Lioba Heimbach}, \bibinfo{person}{Lucianna
  Kiffer}, \bibinfo{person}{Christof~Ferreira Torres}, {and}
  \bibinfo{person}{Roger Wattenhofer}.} \bibinfo{year}{2023}\natexlab{}.
\newblock \showarticletitle{Ethereum's Proposer-Builder Separation: Promises
  and Realities}. In \bibinfo{booktitle}{\emph{Proceedings of the 2023 {ACM} on
  Internet Measurement Conference, {IMC} 2023, Montreal, QC, Canada, October
  24-26, 2023}}, \bibfield{editor}{\bibinfo{person}{Marie{-}Jos{\'{e}}
  Montpetit}, \bibinfo{person}{Aris Leivadeas}, \bibinfo{person}{Steve Uhlig},
  {and} \bibinfo{person}{Mobin Javed}} (Eds.). \bibinfo{publisher}{{ACM}},
  \bibinfo{pages}{406--420}.
\newblock


\bibitem[Heimbach and Wattenhofer(2022a)]%
        {HeimbachW22}
\bibfield{author}{\bibinfo{person}{Lioba Heimbach} {and} \bibinfo{person}{Roger
  Wattenhofer}.} \bibinfo{year}{2022}\natexlab{a}.
\newblock \showarticletitle{Eliminating Sandwich Attacks with the Help of Game
  Theory}. In \bibinfo{booktitle}{\emph{{ASIA} {CCS} '22: {ACM} Asia Conference
  on Computer and Communications Security, Nagasaki, Japan, 30 May 2022 - 3
  June 2022}}, \bibfield{editor}{\bibinfo{person}{Yuji Suga},
  \bibinfo{person}{Kouichi Sakurai}, \bibinfo{person}{Xuhua Ding}, {and}
  \bibinfo{person}{Kazue Sako}} (Eds.). \bibinfo{publisher}{{ACM}},
  \bibinfo{pages}{153--167}.
\newblock


\bibitem[Heimbach and Wattenhofer(2022b)]%
        {heimbach2022eliminating}
\bibfield{author}{\bibinfo{person}{Lioba Heimbach} {and} \bibinfo{person}{Roger
  Wattenhofer}.} \bibinfo{year}{2022}\natexlab{b}.
\newblock \showarticletitle{Eliminating Sandwich Attacks with the Help of Game
  Theory}. In \bibinfo{booktitle}{\emph{{ASIA} {CCS} '22: {ACM} Asia Conference
  on Computer and Communications Security, Nagasaki, Japan, 30 May 2022 - 3
  June 2022}}, \bibfield{editor}{\bibinfo{person}{Yuji Suga},
  \bibinfo{person}{Kouichi Sakurai}, \bibinfo{person}{Xuhua Ding}, {and}
  \bibinfo{person}{Kazue Sako}} (Eds.). \bibinfo{publisher}{{ACM}},
  \bibinfo{pages}{153--167}.
\newblock


\bibitem[Kavousi et~al\mbox{.}(2023)]%
        {kavousi2023blindperm}
\bibfield{author}{\bibinfo{person}{Alireza Kavousi}, \bibinfo{person}{Duc~V
  Le}, \bibinfo{person}{Philipp Jovanovic}, {and} \bibinfo{person}{George
  Danezis}.} \bibinfo{year}{2023}\natexlab{}.
\newblock \showarticletitle{Blindperm: Efficient mev mitigation with an
  encrypted mempool and permutation}.
\newblock \bibinfo{journal}{\emph{Cryptology ePrint Archive}}
  (\bibinfo{year}{2023}).
\newblock


\bibitem[Kokoris-Kogias et~al\mbox{.}(2018)]%
        {kokoris2018omniledger}
\bibfield{author}{\bibinfo{person}{Eleftherios Kokoris-Kogias},
  \bibinfo{person}{Philipp Jovanovic}, \bibinfo{person}{Linus Gasser},
  \bibinfo{person}{Nicolas Gailly}, \bibinfo{person}{Ewa Syta}, {and}
  \bibinfo{person}{Bryan Ford}.} \bibinfo{year}{2018}\natexlab{}.
\newblock \showarticletitle{Omniledger: A secure, scale-out, decentralized
  ledger via sharding}. In \bibinfo{booktitle}{\emph{2018 IEEE symposium on
  security and privacy (SP)}}. IEEE, \bibinfo{pages}{583--598}.
\newblock


\bibitem[L2BEAT(2024)]%
        {l2beat}
\bibfield{author}{\bibinfo{person}{L2BEAT}.} \bibinfo{year}{2024}\natexlab{}.
\newblock \bibinfo{title}{L2BEAT – The state of the layer two ecosystem}.
\newblock
\newblock
\urldef\tempurl%
\url{https://l2beat.com/scaling/summary}
\showURL{%
\tempurl}
\newblock
\shownote{Online; accessed 20 April 2024}.


\bibitem[L2Fees.info(2024)]%
        {l2fees}
\bibfield{author}{\bibinfo{person}{L2Fees.info}.}
  \bibinfo{year}{2024}\natexlab{}.
\newblock \bibinfo{title}{L2 Fees}.
\newblock
\newblock
\urldef\tempurl%
\url{https://l2fees.info/}
\showURL{%
\tempurl}
\newblock
\shownote{Online; accessed 20 April 2024}.


\bibitem[Lewis(2014)]%
        {flashboys_lewis}
\bibfield{author}{\bibinfo{person}{Michael Lewis}.}
  \bibinfo{year}{2014}\natexlab{}.
\newblock \bibinfo{booktitle}{\emph{Flash Boys}}.
\newblock \bibinfo{publisher}{W.W. Norton \& Company}.
\newblock
\showISBNx{978-0-393-24467-0}


\bibitem[Li et~al\mbox{.}(2023)]%
        {li2023demystifying}
\bibfield{author}{\bibinfo{person}{Zihao Li}, \bibinfo{person}{Jianfeng Li},
  \bibinfo{person}{Zheyuan He}, \bibinfo{person}{Xiapu Luo},
  \bibinfo{person}{Ting Wang}, \bibinfo{person}{Xiaoze Ni},
  \bibinfo{person}{Wenwu Yang}, \bibinfo{person}{Xi Chen}, {and}
  \bibinfo{person}{Ting Chen}.} \bibinfo{year}{2023}\natexlab{}.
\newblock \showarticletitle{Demystifying DeFi MEV Activities in Flashbots
  Bundle}. In \bibinfo{booktitle}{\emph{Proceedings of the 2023 ACM SIGSAC
  Conference on Computer and Communications Security}}.
  \bibinfo{pages}{165--179}.
\newblock


\bibitem[McLaughlin et~al\mbox{.}(2023)]%
        {McLaughlinKV23}
\bibfield{author}{\bibinfo{person}{Robert McLaughlin},
  \bibinfo{person}{Christopher Kruegel}, {and} \bibinfo{person}{Giovanni
  Vigna}.} \bibinfo{year}{2023}\natexlab{}.
\newblock \showarticletitle{A Large Scale Study of the Ethereum Arbitrage
  Ecosystem}. In \bibinfo{booktitle}{\emph{32nd {USENIX} Security Symposium,
  {USENIX} Security 2023, Anaheim, CA, USA, August 9-11, 2023}},
  \bibfield{editor}{\bibinfo{person}{Joseph~A. Calandrino} {and}
  \bibinfo{person}{Carmela Troncoso}} (Eds.). \bibinfo{publisher}{{USENIX}
  Association}, \bibinfo{pages}{3295--3312}.
\newblock


\bibitem[Network(2024)]%
        {raiden}
\bibfield{author}{\bibinfo{person}{Raiden Network}.}
  \bibinfo{year}{2024}\natexlab{}.
\newblock \bibinfo{title}{Raiden Network}.
\newblock
\newblock
\urldef\tempurl%
\url{https://raiden.network/}
\showURL{%
\tempurl}
\newblock
\shownote{Online; accessed 20 April 2024}.


\bibitem[Obadia et~al\mbox{.}(2021)]%
        {obadia2021unity}
\bibfield{author}{\bibinfo{person}{Alexandre Obadia}, \bibinfo{person}{Alejo
  Salles}, \bibinfo{person}{Lakshman Sankar}, \bibinfo{person}{Tarun Chitra},
  \bibinfo{person}{Vaibhav Chellani}, {and} \bibinfo{person}{Philip Daian}.}
  \bibinfo{year}{2021}\natexlab{}.
\newblock \showarticletitle{Unity is Strength: {A} Formalization of
  Cross-Domain Maximal Extractable Value}.
\newblock \bibinfo{journal}{\emph{CoRR}}  \bibinfo{volume}{abs/2112.01472}
  (\bibinfo{year}{2021}).
\newblock
\showeprint[arXiv]{2112.01472}
\urldef\tempurl%
\url{https://arxiv.org/abs/2112.01472}
\showURL{%
\tempurl}


\bibitem[OpenZeppelin(2024)]%
        {smart_contract_proxies}
\bibfield{author}{\bibinfo{person}{OpenZeppelin}.}
  \bibinfo{year}{2024}\natexlab{}.
\newblock \bibinfo{title}{Proxy Upgrade Pattern - OpenZeppelin Docs}.
\newblock
\newblock
\urldef\tempurl%
\url{https://docs.openzeppelin.com/upgrades-plugins/1.x/proxies}
\showURL{%
\tempurl}
\newblock
\shownote{Online; accessed 20 April 2024}.


\bibitem[Optimism(2024a)]%
        {optimism}
\bibfield{author}{\bibinfo{person}{Optimism}.}
  \bibinfo{year}{2024}\natexlab{a}.
\newblock \bibinfo{title}{Optimism | Home}.
\newblock
\newblock
\urldef\tempurl%
\url{https://www.optimism.io/}
\showURL{%
\tempurl}
\newblock
\shownote{Online; accessed 20 April 2024}.


\bibitem[Optimism(2024b)]%
        {superchain}
\bibfield{author}{\bibinfo{person}{Optimism}.}
  \bibinfo{year}{2024}\natexlab{b}.
\newblock \bibinfo{title}{Superchain Ecosystem}.
\newblock
\newblock
\urldef\tempurl%
\url{https://www.superchain.eco}
\showURL{%
\tempurl}
\newblock
\shownote{Online; accessed 5 September 2024}.


\bibitem[{\"O}z et~al\mbox{.}(2024)]%
        {oz2024playing}
\bibfield{author}{\bibinfo{person}{Burak {\"O}z}, \bibinfo{person}{Jonas
  Gebele}, \bibinfo{person}{Parshant Singh}, \bibinfo{person}{Filip Rezabek},
  {and} \bibinfo{person}{Florian Matthes}.} \bibinfo{year}{2024}\natexlab{}.
\newblock \showarticletitle{Playing the MEV Game on a First-Come-First-Served
  Blockchain}.
\newblock \bibinfo{journal}{\emph{arXiv preprint arXiv:2401.07992}}
  (\bibinfo{year}{2024}).
\newblock


\bibitem[Perun(2024)]%
        {perun}
\bibfield{author}{\bibinfo{person}{Perun}.} \bibinfo{year}{2024}\natexlab{}.
\newblock \bibinfo{title}{Perun | Blockchains in real-time}.
\newblock
\newblock
\urldef\tempurl%
\url{https://perun.network/}
\showURL{%
\tempurl}
\newblock
\shownote{Online; accessed 20 April 2024}.


\bibitem[Piet et~al\mbox{.}(2022)]%
        {Piet_Fairoze_Weaver_2022}
\bibfield{author}{\bibinfo{person}{Julien Piet}, \bibinfo{person}{Jaiden
  Fairoze}, {and} \bibinfo{person}{Nicholas Weaver}.}
  \bibinfo{year}{2022}\natexlab{}.
\newblock \showarticletitle{Extracting Godl [sic] from the Salt Mines: Ethereum
  Miners Extracting Value}.
\newblock \bibinfo{journal}{\emph{arXiv:2203.15930 [cs]}}
  (\bibinfo{date}{March} \bibinfo{year}{2022}).
\newblock
\urldef\tempurl%
\url{http://arxiv.org/abs/2203.15930}
\showURL{%
\tempurl}
\newblock
\shownote{arXiv: 2203.15930}.


\bibitem[Polygon(2024)]%
        {polygon}
\bibfield{author}{\bibinfo{person}{Polygon}.} \bibinfo{year}{2024}\natexlab{}.
\newblock \bibinfo{title}{Web3, Aggregated.}
\newblock
\newblock
\urldef\tempurl%
\url{https://polygon.technology/}
\showURL{%
\tempurl}
\newblock
\shownote{Online; accessed 20 April 2024}.


\bibitem[Poon and Dryja(2016)]%
        {poon2016bitcoin}
\bibfield{author}{\bibinfo{person}{Joseph Poon} {and} \bibinfo{person}{Thaddeus
  Dryja}.} \bibinfo{year}{2016}\natexlab{}.
\newblock \bibinfo{title}{The bitcoin lightning network: Scalable off-chain
  instant payments}.
\newblock
\newblock


\bibitem[Protocol(2024a)]%
        {across_protocol}
\bibfield{author}{\bibinfo{person}{Across Protocol}.}
  \bibinfo{year}{2024}\natexlab{a}.
\newblock \bibinfo{title}{Home | Across Protocol}.
\newblock
\newblock
\urldef\tempurl%
\url{https://across.to/}
\showURL{%
\tempurl}
\newblock
\shownote{Online; accessed 20 April 2024}.


\bibitem[Protocol(2024b)]%
        {hop_protocol}
\bibfield{author}{\bibinfo{person}{Hop Protocol}.}
  \bibinfo{year}{2024}\natexlab{b}.
\newblock \bibinfo{title}{A Short Explainer | User Docs | Hop Docs}.
\newblock
\newblock
\urldef\tempurl%
\url{https://docs.hop.exchange/basics/a-short-explainer}
\showURL{%
\tempurl}
\newblock
\shownote{Online; accessed 20 April 2024}.


\bibitem[Qin et~al\mbox{.}(2023)]%
        {qin2023blockchain}
\bibfield{author}{\bibinfo{person}{Kaihua Qin}, \bibinfo{person}{Stefanos
  Chaliasos}, \bibinfo{person}{Liyi Zhou}, \bibinfo{person}{Benjamin Livshits},
  \bibinfo{person}{Dawn Song}, {and} \bibinfo{person}{Arthur Gervais}.}
  \bibinfo{year}{2023}\natexlab{}.
\newblock \showarticletitle{The Blockchain Imitation Game}. In
  \bibinfo{booktitle}{\emph{32nd {USENIX} Security Symposium, {USENIX} Security
  2023, Anaheim, CA, USA, August 9-11, 2023}},
  \bibfield{editor}{\bibinfo{person}{Joseph~A. Calandrino} {and}
  \bibinfo{person}{Carmela Troncoso}} (Eds.). \bibinfo{publisher}{{USENIX}
  Association}, \bibinfo{pages}{3961--3978}.
\newblock


\bibitem[Qin et~al\mbox{.}(2022)]%
        {Qin_Zhou_Gervais_2022}
\bibfield{author}{\bibinfo{person}{Kaihua Qin}, \bibinfo{person}{Liyi Zhou},
  {and} \bibinfo{person}{Arthur Gervais}.} \bibinfo{year}{2022}\natexlab{}.
\newblock \showarticletitle{Quantifying Blockchain Extractable Value: How dark
  is the forest?}. In \bibinfo{booktitle}{\emph{43rd {IEEE} Symposium on
  Security and Privacy, {SP} 2022, San Francisco, CA, USA, May 22-26, 2022}}.
  \bibinfo{publisher}{{IEEE}}, \bibinfo{pages}{198--214}.
\newblock


\bibitem[Solidity(2024)]%
        {solidity_metadata}
\bibfield{author}{\bibinfo{person}{Solidity}.} \bibinfo{year}{2024}\natexlab{}.
\newblock \bibinfo{title}{Contract Metadata - Solidity 0.8.26 documentation}.
\newblock
\newblock
\urldef\tempurl%
\url{https://docs.soliditylang.org/en/latest/metadata.html}
\showURL{%
\tempurl}
\newblock
\shownote{Online; accessed 20 April 2024}.


\bibitem[SushiSwap(2024)]%
        {sushiswap}
\bibfield{author}{\bibinfo{person}{SushiSwap}.}
  \bibinfo{year}{2024}\natexlab{}.
\newblock \bibinfo{title}{Buy and Sell Instantly on Sushi.}
\newblock
\newblock
\urldef\tempurl%
\url{https://www.sushi.com/}
\showURL{%
\tempurl}
\newblock
\shownote{Online; accessed 20 April 2024}.


\bibitem[Torres et~al\mbox{.}(2021)]%
        {ferreira2021frontrunner}
\bibfield{author}{\bibinfo{person}{Christof~Ferreira Torres},
  \bibinfo{person}{Ramiro Camino}, {and} \bibinfo{person}{Radu State}.}
  \bibinfo{year}{2021}\natexlab{}.
\newblock \showarticletitle{Frontrunner Jones and the Raiders of the Dark
  Forest: An Empirical Study of Frontrunning on the Ethereum Blockchain}. In
  \bibinfo{booktitle}{\emph{USENIX Security Symposium, Virtual 11-13 August
  2021}}.
\newblock


\bibitem[Tumas et~al\mbox{.}(2023)]%
        {tumas2023ripple}
\bibfield{author}{\bibinfo{person}{Vytautas Tumas}, \bibinfo{person}{Beltran
  Borja~Fiz Pontiveros}, \bibinfo{person}{Christof~Ferreira Torres}, {and}
  \bibinfo{person}{Radu State}.} \bibinfo{year}{2023}\natexlab{}.
\newblock \showarticletitle{A Ripple for Change: Analysis of Frontrunning in
  the XRP Ledger}. In \bibinfo{booktitle}{\emph{2023 IEEE International
  Conference on Blockchain and Cryptocurrency (ICBC)}}. IEEE,
  \bibinfo{pages}{1--9}.
\newblock


\bibitem[Uniswap(2024)]%
        {uniswap}
\bibfield{author}{\bibinfo{person}{Uniswap}.} \bibinfo{year}{2024}\natexlab{}.
\newblock \bibinfo{title}{Uniswap Protocol}.
\newblock
\newblock
\urldef\tempurl%
\url{https://uniswap.org/}
\showURL{%
\tempurl}
\newblock
\shownote{Online; accessed 20 April 2024}.


\bibitem[Wadhwa et~al\mbox{.}(2023)]%
        {wadhwa2023data}
\bibfield{author}{\bibinfo{person}{Sarisht Wadhwa}, \bibinfo{person}{Luca
  Zanolini}, \bibinfo{person}{Francesco D'Amato}, \bibinfo{person}{Aditya
  Asgaonkar}, \bibinfo{person}{Chengrui Fang}, \bibinfo{person}{Fan Zhang},
  {and} \bibinfo{person}{Kartik Nayak}.} \bibinfo{year}{2023}\natexlab{}.
\newblock \showarticletitle{Data Independent Order Policy Enforcement:
  Limitations and Solutions}.
\newblock \bibinfo{journal}{\emph{Cryptology ePrint Archive}}
  (\bibinfo{year}{2023}).
\newblock


\bibitem[Wang et~al\mbox{.}(2021)]%
        {wang2021towards}
\bibfield{author}{\bibinfo{person}{Dabao Wang}, \bibinfo{person}{Siwei Wu},
  \bibinfo{person}{Ziling Lin}, \bibinfo{person}{Lei Wu},
  \bibinfo{person}{Xingliang Yuan}, \bibinfo{person}{Yajin Zhou},
  \bibinfo{person}{Haoyu Wang}, {and} \bibinfo{person}{Kui Ren}.}
  \bibinfo{year}{2021}\natexlab{}.
\newblock \showarticletitle{Towards a first step to understand flash loan and
  its applications in defi ecosystem}. In \bibinfo{booktitle}{\emph{Proceedings
  of the Ninth International Workshop on Security in Blockchain and Cloud
  Computing}}. \bibinfo{pages}{23--28}.
\newblock


\bibitem[Wang et~al\mbox{.}(2022)]%
        {wang2022cyclic}
\bibfield{author}{\bibinfo{person}{Ye Wang}, \bibinfo{person}{Yan Chen},
  \bibinfo{person}{Haotian Wu}, \bibinfo{person}{Liyi Zhou},
  \bibinfo{person}{Shuiguang Deng}, {and} \bibinfo{person}{Roger Wattenhofer}.}
  \bibinfo{year}{2022}\natexlab{}.
\newblock \showarticletitle{Cyclic arbitrage in decentralized exchanges}. In
  \bibinfo{booktitle}{\emph{Companion Proceedings of the Web Conference 2022}}.
  \bibinfo{pages}{12--19}.
\newblock


\bibitem[Weintraub et~al\mbox{.}(2022)]%
        {flashbotinthepan}
\bibfield{author}{\bibinfo{person}{Ben Weintraub}, \bibinfo{person}{Christof
  Ferreira~Torres}, \bibinfo{person}{Cristina Nita-Rotaru}, {and}
  \bibinfo{person}{Radu State}.} \bibinfo{year}{2022}\natexlab{}.
\newblock \showarticletitle{A Flash(bot) in the Pan: Measuring Maximal
  Extractable Value in Private Pools}. In \bibinfo{booktitle}{\emph{Proceedings
  of the 22nd ACM Internet Measurement Conference (IMC ’22)}}.
  \bibinfo{publisher}{Association for Computing Machinery},
  \bibinfo{address}{Nice, France}.
\newblock
\showISBNx{978-1-4503-9259-4}


\bibitem[Wood et~al\mbox{.}(2014)]%
        {ethereum}
\bibfield{author}{\bibinfo{person}{Gavin Wood} {et~al\mbox{.}}}
  \bibinfo{year}{2014}\natexlab{}.
\newblock \showarticletitle{Ethereum: A secure decentralised generalised
  transaction ledger}.
\newblock \bibinfo{journal}{\emph{Ethereum project yellow paper}}
  \bibinfo{volume}{151}, \bibinfo{number}{2014} (\bibinfo{year}{2014}),
  \bibinfo{pages}{1--32}.
\newblock


\bibitem[Wormhole(2024)]%
        {wormhole}
\bibfield{author}{\bibinfo{person}{Wormhole}.} \bibinfo{year}{2024}\natexlab{}.
\newblock \bibinfo{title}{The best way to build cross-chain}.
\newblock
\newblock
\urldef\tempurl%
\url{https://wormhole.com/}
\showURL{%
\tempurl}
\newblock
\shownote{Online; accessed 20 April 2024}.


\bibitem[Yu et~al\mbox{.}(2020)]%
        {yu2020survey}
\bibfield{author}{\bibinfo{person}{Guangsheng Yu}, \bibinfo{person}{Xu Wang},
  \bibinfo{person}{Kan Yu}, \bibinfo{person}{Wei Ni}, \bibinfo{person}{J~Andrew
  Zhang}, {and} \bibinfo{person}{Ren~Ping Liu}.}
  \bibinfo{year}{2020}\natexlab{}.
\newblock \showarticletitle{Survey: Sharding in blockchains}.
\newblock \bibinfo{journal}{\emph{IEEE Access}}  \bibinfo{volume}{8}
  (\bibinfo{year}{2020}), \bibinfo{pages}{14155--14181}.
\newblock


\bibitem[Zhang et~al\mbox{.}(2023a)]%
        {zhang2023front}
\bibfield{author}{\bibinfo{person}{Jianting Zhang}, \bibinfo{person}{Wuhui
  Chen}, \bibinfo{person}{Sifu Luo}, \bibinfo{person}{Tiantian Gong},
  \bibinfo{person}{Zicong Hong}, {and} \bibinfo{person}{Aniket Kate}.}
  \bibinfo{year}{2023}\natexlab{a}.
\newblock \showarticletitle{Front-running Attack in Distributed Sharded Ledgers
  and Fair Cross-shard Consensus}.
\newblock \bibinfo{journal}{\emph{arXiv preprint arXiv:2306.06299}}
  (\bibinfo{year}{2023}).
\newblock


\bibitem[Zhang et~al\mbox{.}(2023b)]%
        {exploitZhang2023}
\bibfield{author}{\bibinfo{person}{Zhuo Zhang}, \bibinfo{person}{Zhiqiang Lin},
  \bibinfo{person}{Marcelo Morales}, \bibinfo{person}{Xiangyu Zhang}, {and}
  \bibinfo{person}{Kaiyuan Zhang}.} \bibinfo{year}{2023}\natexlab{b}.
\newblock \showarticletitle{Your Exploit is Mine: Instantly Synthesizing
  Counterattack Smart Contract}. In \bibinfo{booktitle}{\emph{32nd {USENIX}
  Security Symposium, {USENIX} Security 2023, Anaheim, CA, USA, August 9-11,
  2023}}, \bibfield{editor}{\bibinfo{person}{Joseph~A. Calandrino} {and}
  \bibinfo{person}{Carmela Troncoso}} (Eds.). \bibinfo{publisher}{{USENIX}
  Association}, \bibinfo{pages}{1757--1774}.
\newblock


\bibitem[Zhou et~al\mbox{.}(2021)]%
        {zhou2021a2mm}
\bibfield{author}{\bibinfo{person}{Liyi Zhou}, \bibinfo{person}{Kaihua Qin},
  {and} \bibinfo{person}{Arthur Gervais}.} \bibinfo{year}{2021}\natexlab{}.
\newblock \showarticletitle{{A2MM:} Mitigating Frontrunning, Transaction
  Reordering and Consensus Instability in Decentralized Exchanges}.
\newblock \bibinfo{journal}{\emph{CoRR}}  \bibinfo{volume}{abs/2106.07371}
  (\bibinfo{year}{2021}).
\newblock
\showeprint[arXiv]{2106.07371}
\urldef\tempurl%
\url{https://arxiv.org/abs/2106.07371}
\showURL{%
\tempurl}


\bibitem[zkSync(2024a)]%
        {zkevm}
\bibfield{author}{\bibinfo{person}{zkSync}.} \bibinfo{year}{2024}\natexlab{a}.
\newblock \bibinfo{title}{zkEVM FaQ | zkSync Documentaion}.
\newblock
\newblock
\urldef\tempurl%
\url{https://docs.zksync.io/zkevm/}
\showURL{%
\tempurl}
\newblock
\shownote{Online; accessed 20 April 2024}.


\bibitem[zkSync(2024b)]%
        {zksync}
\bibfield{author}{\bibinfo{person}{zkSync}.} \bibinfo{year}{2024}\natexlab{b}.
\newblock \bibinfo{title}{zkSync | Scaling the Ethos and technology of
  Ethereum}.
\newblock
\newblock
\urldef\tempurl%
\url{https://zksync.io/}
\showURL{%
\tempurl}
\newblock
\shownote{Online; accessed 20 April 2024}.


\bibitem[zkSync(2024c)]%
        {zkcompiler}
\bibfield{author}{\bibinfo{person}{zkSync}.} \bibinfo{year}{2024}\natexlab{c}.
\newblock \bibinfo{title}{zkSync Era Developer Tools | Compiler Toolchain |
  Overview}.
\newblock
\newblock
\urldef\tempurl%
\url{https://era.zksync.io/docs/tools/compiler-toolchain/overview.html}
\showURL{%
\tempurl}
\newblock
\shownote{Online; accessed 20 April 2024}.


\bibitem[zkSync Era~Explorer(2024)]%
        {zksync_explorer_api}
\bibfield{author}{\bibinfo{person}{zkSync Era~Explorer}.}
  \bibinfo{year}{2024}\natexlab{}.
\newblock \bibinfo{title}{ZkSync Block Explorer API}.
\newblock
\newblock
\urldef\tempurl%
\url{https://block-explorer-api.mainnet.zksync.io/docs}
\showURL{%
\tempurl}
\newblock
\shownote{Online; accessed 20 April 2024}.


\end{thebibliography}

\begin{table*}
    \centering
    \begin{adjustbox}{width=\textwidth}
    \begin{tabular}{l l l c}
        \toprule
        \textbf{Category} & \textbf{Protocol/Chain} & \textbf{Event Name} & \textbf{Event Topic Hash} \\
        \midrule
        \multirow{6}{*}{Arbitrage} & Uniswap V2 & \textit{Swap} & \texttt{0xd78ad95fa46c994b6551d0da85fc275fe613ce37657fb8d5e3d130840159d822} \\
        & Uniswap V3 & \textit{Swap} & \texttt{0xc42079f94a6350d7e6235f29174924f928cc2ac818eb64fed8004e115fbcca67} \\
        & Balancer V1 & \textit{LOG\_SWAP} & \texttt{0x908fb5ee8f16c6bc9bc3690973819f32a4d4b10188134543c88706e0e1d43378} \\
        & Balancer V2 & \textit{Swap} & \texttt{0x2170c741c41531aec20e7c107c24eecfdd15e69c9bb0a8dd37b1840b9e0b207b} \\
        & Curve & \textit{TokenExchangeUnderlying} & \texttt{0xd013ca23e77a65003c2c659c5442c00c805371b7fc1ebd4c206c41d1536bd90b} \\
        & Curve & \textit{TokenExchange} & \texttt{0x8b3e96f2b889fa771c53c981b40daf005f63f637f1869f707052d15a3dd97140} \\
        \midrule
        \multirow{5}{*}{Liquidations} &
        Aave V1 & \textit{LiquidationCall} & \texttt{0x56864757fd5b1fc9f38f5f3a981cd8ae512ce41b902cf73fc506ee369c6bc237} \\ 
        & Aave V2/V3 & \textit{LiquidationCall} & \texttt{0xe413a321e8681d831f4dbccbca790d2952b56f977908e45be37335533e005286} \\ 
        & Compound V2 & \textit{LiquidateBorrow} & \texttt{0x298637f684da70674f26509b10f07ec2fbc77a335ab1e7d6215a4b2484d8bb52} \\
        & Compound & \textit{Redeem} & \texttt{0xe5b754fb1abb7f01b499791d0b820ae3b6af3424ac1c59768edb53f4ec31a929} \\
        & Compound & \textit{Redeem} & \texttt{0xe02f6383e19e87c24e0c03e2cd5dbd05156cb29a1b0f3dbca1fa3430e444f63d} \\
        \midrule
        \multirow{1}{*}{Sandwiches} & ERC-20 & \textit{Transfer} & \texttt{0xddf252ad1be2c89b69c2b068fc378daa952ba7f163c4a11628f55a4df523b3ef} \\
        \midrule
        \multirow{1}{*}{Oracle Updates} &
        Chainlink & \textit{AnswerUpdated} & \texttt{0x0559884fd3a460db3073b7fc896cc77986f16e378210ded43186175bf646fc5f} \\
        \midrule
        \multirow{4}{*}{Flash Loans} &
        Aave V1 & \textit{FlashLoan} & \texttt{0x5b8f46461c1dd69fb968f1a003acee221ea3e19540e350233b612ddb43433b55} \\
        & Aave V2 & \textit{FlashLoan} & \texttt{0x631042c832b07452973831137f2d73e395028b44b250dedc5abb0ee766e168ac} \\
        & Aave V3 & \textit{FlashLoan} & \texttt{0xefefaba5e921573100900a3ad9cf29f222d995fb3b6045797eaea7521bd8d6f0} \\
        & Balancer & \textit{FlashLoan} & \texttt{0x0d7d75e01ab95780d3cd1c8ec0dd6c2ce19e3a20427eec8bf53283b6fb8e95f0} \\
        \midrule
        \multirow{4}{*}{L1 Messages} &
        Arbitrum & \textit{InboxMessageDelivered} & \texttt{0xff64905f73a67fb594e0f940a8075a860db489ad991e032f48c81123eb52d60b} \\
        & Optimism & \textit{TransactionEnqueued} & \texttt{0x4b388aecf9fa6cc92253704e5975a6129a4f735bdbd99567df4ed0094ee4ceb5} \\
        & Optimism & \textit{TransactionDeposited} & \texttt{0xb3813568d9991fc951961fcb4c784893574240a28925604d09fc577c55bb7c32} \\
        & zkSync & \textit{NewPriorityRequest} & \texttt{0x4531cd5795773d7101c17bdeb9f5ab7f47d7056017506f937083be5d6e77a382} \\
        \midrule
        \multirow{2}{*}{L2 Messages} & 
        Arbitrum & \textit{RedeemScheduled} & \texttt{0x5ccd009502509cf28762c67858994d85b163bb6e451f5e9df7c5e18c9c2e123e} \\
        & Optimism & \textit{RelayedMessage} & \texttt{0x4641df4a962071e12719d8c8c8e5ac7fc4d97b927346a3d7a335b1f7517e133c} \\
        \midrule
        \multirow{3}{*}{Victim Inference} & ERC-20 & \textit{Transfer} & \texttt{0xddf252ad1be2c89b69c2b068fc378daa952ba7f163c4a11628f55a4df523b3ef} \\
        & StableSwap & \textit{TokenSwap} & \texttt{0xc6c1e0630dbe9130cc068028486c0d118ddcea348550819defd5cb8c257f8a38} \\
        & Uniswap V3 & \textit{Swap} & \texttt{0xc42079f94a6350d7e6235f29174924f928cc2ac818eb64fed8004e115fbcca67} \\
        \bottomrule
    \end{tabular}
    \end{adjustbox}
    \caption{Overview of events used in this work to perform measurements across chains and DeFi protocols.}
    \label{tab:events}
\end{table*}

\begin{table*}
    \centering
    \begin{adjustbox}{width=\textwidth}    
    \begin{tabular}{l | r r r r | r r r r | r r r r}
        \toprule
        & \multicolumn{4}{c|}{\textbf{Arbitrage}} & \multicolumn{4}{c|}{\textbf{Liquidation}} & \multicolumn{4}{c}{\textbf{Sandwiching}} \\
        \textbf{DeFi Protocol} & \textbf{Ethereum} & \textbf{Arbitrum} & \textbf{Optimism} & \textbf{zkSync} & \textbf{Ethereum} & \textbf{Arbitrum} & \textbf{Optimism} & \textbf{zkSync} & \textbf{Ethereum} & \textbf{Arbitrum} & \textbf{Optimism} & \textbf{zkSync} \\
        \midrule
        Aave & - & - & - & - & \num{30734} & \num{2829} & \num{3358} & - & - & - & - & - \\
        Balancer V1 & \num{349464} & - & - & - & - & - & - & - & - & - & - & - \\
        Balancer V2 & \num{127866} & \num{167294} & \num{83044} & - & - & - & - & - & - & - & - & - \\
        Compound & - & - & - & - & \num{24038} & \num{2448} & \num{1989} & \num{325} & - & - & - & - \\
        Curve & \num{45315} & \num{27238} & \num{30186} & - & - & - & - & - & - & - & - & - \\
        Uniswap V2 & \num{2572061} & \num{1074085} & \num{990746} & \num{108064} & - & - & - & - & \num{2089610} & - & - & - \\
        Uniswap V3 & \num{1450698} & \num{1534733} & \num{830359} & \num{2890} & - & - & - & - & \num{265843} & - & - & - \\
        \midrule
        \textbf{Total Unique} & \textbf{\num{2901740}} & \textbf{\num{1746083}} & \textbf{\num{1153366}} & \textbf{\num{108070}} & \textbf{\num{54772}} & \textbf{\num{5277}} & \textbf{\num{5347}} & \textbf{\num{325}} & \textbf{\num{2334566}} & \textbf{-} & \textbf{-} & \textbf{-} \\ 
        \bottomrule
    \end{tabular}
    \end{adjustbox}
    \caption{Total number of arbitrages, liquidations, and sandwiches detected across different DeFi protocols and chains.}
    \label{tab:arb_lib_count}
\end{table*}

\appendix

\begin{table*}
    \centering
    \small
    \begin{tabular}{c c c r c l c r}
    \toprule
    \textbf{Rollup} & \textbf{DEX Contract Address} & \multicolumn{1}{c}{\textbf{Protocol}} & \multicolumn{3}{c}{\textbf{Liquidity Pool}} & \textbf{AMM Algorithm} & \multicolumn{1}{c}{\textbf{Swap Txs}} \\
    \midrule 
    \multirow{6}{*}{Arbitrum} & \href{https://arbiscan.io/address/0x652d27c0F72771Ce5C76fd400edD61B406Ac6D97}{\footnotesize{\texttt{0x652d27c0F72771Ce5C76fd400edD61B406Ac6D97}}} & \href{https://hop.exchange/}{Hop Protocol} & ETH & $\leftrightarrow$ & hETH  & StableSwap & 76,312 \\
    & \href{https://arbiscan.io/address/0x10541b07d8Ad2647Dc6cD67abd4c03575dade261}{\footnotesize{\texttt{0x10541b07d8Ad2647Dc6cD67abd4c03575dade261}}} & \href{https://hop.exchange/}{Hop Protocol} & USDC & $\leftrightarrow$ & hUSDC & StableSwap & 8,014 \\
    & \href{https://arbiscan.io/address/0x18f7402B673Ba6Fb5EA4B95768aABb8aaD7ef18a}{\footnotesize{\texttt{0x18f7402B673Ba6Fb5EA4B95768aABb8aaD7ef18a}}} & \href{https://hop.exchange/}{Hop Protocol} & USDT & $\leftrightarrow$ & hUSDT & StableSwap & 2,525 \\
    & \href{https://arbiscan.io/address/0xa5A33aB9063395A90CCbEa2D86a62EcCf27B5742}{\footnotesize{\texttt{0xa5A33aB9063395A90CCbEa2D86a62EcCf27B5742}}} & \href{https://hop.exchange/}{Hop Protocol} & DAI & $\leftrightarrow$ & hDAI & StableSwap & 996 \\
    & \href{https://arbiscan.io/address/0x0Ded0d521AC7B0d312871D18EA4FDE79f03Ee7CA}{\footnotesize{\texttt{0x0Ded0d521AC7B0d312871D18EA4FDE79f03Ee7CA}}} & \href{https://hop.exchange/}{Hop Protocol} & rETH & $\leftrightarrow$ & hrETH & StableSwap & 65 \\
    & \href{https://arbiscan.io/address/0xFFe42d3Ba79Ee5Ee74a999CAd0c60EF1153F0b82}{\footnotesize{\texttt{0xFFe42d3Ba79Ee5Ee74a999CAd0c60EF1153F0b82}}} & \href{https://hop.exchange/}{Hop Protocol} & MAGIC & $\leftrightarrow$ & hMAGIC & StableSwap & 9 \\
    \midrule
    \multirow{9}{*}{Optimism} & \href{https://optimistic.etherscan.io/address/0xaa30D6bba6285d0585722e2440Ff89E23EF68864}{\footnotesize{\texttt{0xaa30D6bba6285d0585722e2440Ff89E23EF68864}}} & \href{https://hop.exchange/}{Hop Protocol} & ETH & $\leftrightarrow$ & hETH & StableSwap & \num{70118} \\
    & \href{https://optimistic.etherscan.io/address/0x3c0FFAca566fCcfD9Cc95139FEF6CBA143795963}{\footnotesize{\texttt{0x3c0FFAca566fCcfD9Cc95139FEF6CBA143795963}}} & \href{https://hop.exchange/}{Hop Protocol} & USDC & $\leftrightarrow$ & hUSDC & StableSwap & \num{8756} \\
    & \href{https://optimistic.etherscan.io/address/0xeC4B41Af04cF917b54AEb6Df58c0f8D78895b5Ef}{\footnotesize{\texttt{0xeC4B41Af04cF917b54AEb6Df58c0f8D78895b5Ef}}} & \href{https://hop.exchange/}{Hop Protocol} & USDT & $\leftrightarrow$ & hUSDT & StableSwap & \num{1696} \\
    & \href{https://optimistic.etherscan.io/address/0xF181eD90D6CfaC84B8073FdEA6D34Aa744B41810}{\footnotesize{\texttt{0xF181eD90D6CfaC84B8073FdEA6D34Aa744B41810}}} & \href{https://hop.exchange/}{Hop Protocol} & DAI & $\leftrightarrow$ & hDAI & StableSwap & \num{1315} \\
    & \href{https://optimistic.etherscan.io/address/0x1990BC6dfe2ef605Bfc08f5A23564dB75642Ad73}{\footnotesize{\texttt{0x1990BC6dfe2ef605Bfc08f5A23564dB75642Ad73}}} & \href{https://hop.exchange/}{Hop Protocol} & SNX & $\leftrightarrow$ & hSNX & StableSwap & \num{698} \\
    & \href{https://optimistic.etherscan.io/address/0x8d4063E82A4Db8CdAed46932E1c71e03CA69Bede}{\footnotesize{\texttt{0x8d4063E82A4Db8CdAed46932E1c71e03CA69Bede}}} & \href{https://hop.exchange/}{Hop Protocol} & sUSD & $\leftrightarrow$ & hsUSD & StableSwap & \num{105} \\
    & \href{https://optimistic.etherscan.io/address/0x9Dd8685463285aD5a94D2c128bda3c5e8a6173c8}{\footnotesize{\texttt{0x9Dd8685463285aD5a94D2c128bda3c5e8a6173c8}}} & \href{https://hop.exchange/}{Hop Protocol} & rETH & $\leftrightarrow$ & hrETH & StableSwap & \num{63} \\
    & \href{https://optimistic.etherscan.io/address/0x6e39aCC0Dd292a70D92c447ebCcB8728f4eD5FE4}{\footnotesize{\texttt{0x6e39aCC0Dd292a70D92c447ebCcB8728f4eD5FE4}}} & \href{https://perp.com/}{Perpetual Protocol} & cCRV & $\leftrightarrow$ & vUSD & Uniswap V3 & 1 \\
    & \href{https://optimistic.etherscan.io/address/0x36B18618c4131D8564A714fb6b4D2B1EdADc0042}{\footnotesize{\texttt{0x36B18618c4131D8564A714fb6b4D2B1EdADc0042}}} & \href{https://perp.com/}{Perpetual Protocol} & vUSD & $\leftrightarrow$ & vETH & Uniswap V3 & 1 \\
    \bottomrule
    \end{tabular}
    \caption{Overview of identified potential victim transactions performing swaps on L2 via L1 transactions.}
    \label{tab:victims}
\end{table*}

\begin{table*}
    \centering
    \begin{adjustbox}{width=\textwidth}
    \begin{tabular}{c l | r r r r r | r r r r r}
        \toprule
        & & \multicolumn{5}{c|}{\textbf{Arbitrum}} & \multicolumn{5}{c}{\textbf{Optimism}} \\
        \multicolumn{2}{c|}{\textbf{Attack Strategy}} &         \textbf{1K-Cap.} & 
        \textbf{10K-Cap.} & 
        \textbf{100K-Cap.} & 
        \textbf{1M-Cap.}& 
        \textbf{$\infty$-Cap.} & 
        \textbf{1K-Cap.} & 
        \textbf{10K-Cap.} & 
        \textbf{100K-Cap.} & 
        \textbf{1M-Cap.}& 
        \textbf{$\infty$-Cap.} \\
        \midrule

        \multirow{6}{*}{$S_1$} 
        & $P_{\text{Total}}$ & \num{38.93} & \num{1235.10} & \num{14631.47} & \num{129656.26} & \num{1224200.05} & \num{226.35} & \num{4966.80} & \num{68991.44} & \num{264374.17} & \num{739695.63} \\
        & $P_{\text{Max}}$ & 24.39 & 175.71 & \num{1310.81} & \num{35570.85} & \num{785862.73} & 108.03 & \num{1533.57} & \num{7481.66} & \num{7481.66} & \num{138184.20} \\
        & $P_{\text{Mean}}$ & 3.89 & 14.36 & 66.51 & 448.64 & \num{3898.73} & 32.34 & 25.34 & 129.20 & 342.00 & 896.60 \\
        & $P_{\text{Median}}$ & 0.71 & 7.07 & 33.27 & 88.44 & 106.07 & 6.33 & 8.30 & 50.30 & 165.32 & 187.80 \\
        & $P_{\text{Min}}$ & 0.00 & 0.00 & 0.00 & 0.00 & 0.00 & 1.31 & 0.07 & 0.41 & 0.41 & 0.41\\
        
        \midrule
        \multirow{6}{*}{$S_2$} 
        & $P_{\text{Total}}$ & 59.93 & \num{1606.38} & \num{15775.96} & \num{131257.43} & \num{1225968.21} & 295.37 & \num{6688.07} & \num{75679.86} & \num{276412.81} & \num{752736.05}\\
        & $P_{\text{Max}}$ & 27.20 & 178.52 & \num{1312.94} & \num{35576.04} & \num{785866.70} & 134.46 & \num{1560.01} & \num{7508.09} & \num{7508.09} & \num{138205.96} \\
        & $P_{\text{Mean}}$ & 3.53 & 12.75 & 66.57 & 444.94 & \num{3807.35} &10.19&22.15&117.70&327.12&838.24 \\
        & $P_{\text{Median}}$ & 1.38 & 6.59 & 34.21 & 93.08 & 108.08&0.89&9.48&43.83&148.25&168.42\\
        & $P_{\text{Min}}$ & 0.04 & 0.01 & 0.21 & 0.45 & 0.45 & 0.02 & 0.01 & 0.04 & 0.04 & 0.04\\
        
        \midrule
        \multirow{6}{*}{$S_3$} 
        & $P_{\text{Total}}$ & 237.47 & \num{2410.52} & \num{17108.52} & \num{132926.89} & \num{1227808.79} & \num{1213.84} & \num{11369.12} & \num{86369.52} & \num{290464.40} & \num{767972.37} \\ 
        & $P_{\text{Max}}$ &30.00 & 181.33 & \num{1315.07} & \num{35581.23} & \num{785870.67} & 160.89 & \num{1586.44} & \num{7534.52} & \num{7534.52} & \num{138227.72} \\ 
        & $P_{\text{Mean}}$ & 1.10 & 9.49 & 61.76 & 415.40 & \num{3558.87} &1.71 & 13.95 & 101.97 & 301.00 & 756.62 \\ 
        & $P_{\text{Median}}$ &0.53 & 4.82 & 27.59 & 84.10 & 101.04 & 0.57 & 4.73 & 37.46 & 119.93 & 134.35 \\ 
        & $P_{\text{Min}}$ &0.00 & 0.04 & 0.04 & 0.04 & 0.04 & 0.00 & 0.00 & 0.04 & 0.04 & 0.04 \\ 
        \bottomrule
    \end{tabular}
    \end{adjustbox}
    \caption{Profit in USD for each attack strategy across Arbitrum and Optimism simulating different types of attacker capital.}
    \label{tab:sandwich_profitability}
\end{table*}

\begin{table*}
    \centering
    \begin{adjustbox}{width=\textwidth}
    \begin{tabular}{c c c c c | c c c | c c c}
        \toprule
        & & \multicolumn{3}{c|}{\textbf{Attack Strategy $S_1$}} & \multicolumn{3}{c|}{\textbf{Attack Strategy $S_2$}} & \multicolumn{3}{c}{\textbf{Attack Strategy $S_3$}} \\
        & & \textbf{Arbitrum} & \textbf{Optimism} & \textbf{zkSync} & \textbf{Arbitrum} & \textbf{Optimism} & \textbf{zkSync} & \textbf{Arbitrum} & \textbf{Optimism} & \textbf{zkSync} \\
        \midrule
        \multirow{2}{*}{ $T_{A_1}$} & L1 & \cellcolor{red!20!white} \href{https://sepolia.etherscan.io/tx/0x72376cbafbe09901f9a027b18c3efc3260c425c37fb8bee806b366397086561d}{\footnotesize{\texttt{0x7237...561d}}} 
        & \cellcolor{red!20!white} \href{https://sepolia.etherscan.io/tx/0x0c96a0bf59950157b65da5b80fc96ab769dd464d4c1376ccd0bc0b86f831e38d}{\footnotesize{\texttt{0x0c96...e38d}}} 
        & \cellcolor{red!20!white} \href{https://sepolia.etherscan.io/tx/0xd27af0131d47ee505a6675ea4372d3215d19bce3ed30da2832a98887b2423148}{\footnotesize{\texttt{0xd27a...3148}}} & 
        \cellcolor{red!20!white} \href{https://sepolia.etherscan.io/tx/0x3c261ff7b6a6a1b83af3bf4b7a50883addf53c488ab9bcfa929056fb54ae22cb}{\footnotesize{\texttt{0x3c26...22cb}}} & \cellcolor{red!20!white} \href{https://sepolia.etherscan.io/tx/0x78004cc692566528a48995228a78e64d9b5586cdc2f0fe7ed3476f57adbd8182}{\footnotesize{\texttt{0x7800...8182}}} & \cellcolor{red!20!white} \href{https://sepolia.etherscan.io/tx/0xe561502b7203a4151d822241a6de2c7873896888a3a71e6a78c3e80c49c5f522}{\footnotesize{\texttt{0xe561...f522}}} & 
        - & - & - \\
         & L2 & 
         \cellcolor{red!20!white} \href{https://sepolia.arbiscan.io/tx/0xe5d74adb9d534d47db4e8d8efd44af35fd07e03bf73b068fc4d4e8e9eb108c28}{\footnotesize{\texttt{0xe5d7...8c28}}}& 
         \cellcolor{red!20!white} \href{https://sepolia-optimism.etherscan.io/tx/0x2f266df801ff3b71dc595f0e4970c97f73396485a93aaec464afddf9e8efc4f8}{\footnotesize{\texttt{0x2f26...c4f8}}}  & 
         \cellcolor{red!20!white} \href{https://sepolia-era.zksync.network/tx/0xefe7c2c081c96557adb4ff5f41ebf889a73252e8471b287a52e4d9254bce4ed3}{\footnotesize{\texttt{0xefe7...4ed3}}} &
         \cellcolor{red!20!white} \href{https://sepolia.arbiscan.io/tx/0xb5bb1eea25224e861513c36c805f2b526a18bf1426075dcbc1fa89e4d81cfc1a}{\footnotesize{\texttt{0xb5bb...fc1a}}} & 
         \cellcolor{red!20!white} \href{https://sepolia-optimism.etherscan.io/tx/0x059daf9e6a8d051cbe9a2fbfb1b9d50b4f56333d0b82838af124a0fe61e5f362}{\footnotesize{\texttt{0x059d...f362}}}  & 
         \cellcolor{red!20!white} \href{https://sepolia-era.zksync.network/tx/0x72758c56eda7939ab498b147f55feb00666b63c9657cc509447031009cf17e92}{\footnotesize{\texttt{0x7275...7e92}}} &
         \cellcolor{red!20!white} \href{https://sepolia.arbiscan.io/tx/0xc9c049760b3326635ccebda9eac2a6ef3cd7b81c59bc0a393f2e658f90e63321}{\footnotesize{\texttt{0xc9c0...3321}}} & 
         \cellcolor{red!20!white} \href{https://sepolia-optimism.etherscan.io/tx/0x4158d2c20508c4bf03fbc9b17aec493b2fb617ea03228f362201b82dc4906c16}{\footnotesize{\texttt{0x4158...6c16}}} & 
         \cellcolor{red!20!white} \href{https://sepolia-era.zksync.network/tx/0xb53ce9ccea03a6fbe61fad6d47c7f9574fee397948f2817b5bd747e3693db39a}{\footnotesize{\texttt{0xb53c...b39a}}} 
         \\ \cline{2-11}
         \multirow{2}{*}{$T_V$} 
         & L1 & 
         \cellcolor{blue!20!white} \href{https://sepolia.etherscan.io/tx/0x875f096dd6dfb509f43f232814fd8f7b34c6093d905fb464f9ffd65b79e5ab89}{\footnotesize{\texttt{0x875f...ab89}}} & \cellcolor{blue!20!white} \href{https://sepolia.etherscan.io/tx/0x3bb4c397fc3172a3628cf0b395f160796cdb8c0ca1f2d3a9c0676b211a1f1dda}{\footnotesize{\texttt{0x3bb4...1dda}}}  & \cellcolor{blue!20!white} \href{https://sepolia.etherscan.io/tx/0xaef7f4ed9fefea8e9a0769b3b41f02f8ed4733ef4c1829c8cc099e0ad09c3ba8}{\footnotesize{\texttt{0xaef7...3ba8}}} &
         \cellcolor{blue!20!white} \href{https://sepolia.etherscan.io/tx/0x55488bfd38434cd23aecab767b70a042de203a2d6497eafe15db1eaab3bce448}{\footnotesize{\texttt{0x5548...e448}}} & \cellcolor{blue!20!white} \href{https://sepolia.etherscan.io/tx/0x4bdaf3191e24ed8c4628a409ffddb9a64deda9cec3b0ff028eda134163aef0e0}{\footnotesize{\texttt{0x4bda...f0e0}}} & \cellcolor{blue!20!white} \href{https://sepolia.etherscan.io/tx/0xba2f7129e00dce304753df8c2736d4adcb465fcaf65a92370a939f5967c9c173}{\footnotesize{\texttt{0xba2f...c173}}} &
         \cellcolor{blue!20!white} \href{https://sepolia.etherscan.io/tx/0x6b100c05f8a3fe48625002a8e2159d83955e2b05b3036c1b68b01c418a13e828}{\footnotesize{\texttt{0x6b10...e828}}} & \cellcolor{blue!20!white} \href{https://sepolia.etherscan.io/tx/0xe843550c3423f2a27ced281c5dda32805e0e0519bce942efec0593870fad219d}{\footnotesize{\texttt{0xe843...219d}}} & \cellcolor{blue!20!white} \href{https://sepolia.etherscan.io/tx/0x015981d3111d2126d01f45bd065339501b3a8fea758332336f6ac99d5035b270}{\footnotesize{\texttt{0x0159...b270}}}
         \\ 
         & L2 & 
         \cellcolor{blue!20!white} \href{https://sepolia.arbiscan.io/tx/0x7426864411b565bb904157267cd1469f017aff50c377c10b0c3717885c5948b5}{\footnotesize{\texttt{0x7426...48b5}}} & 
         \cellcolor{blue!20!white} \href{https://sepolia-optimism.etherscan.io/tx/0x0169f3424944a98f4da8e61da4d6bc2823a6b5e16b56f795d49fc913beccb82c}{\footnotesize{\texttt{0x0169...b82c}}} & 
         \cellcolor{blue!20!white} \href{https://sepolia-era.zksync.network/tx/0xcc7c5cd049d5aed3571d1da8eabad5bf19a078829f495fcb8a0f3d1e2bcfe78e}{\footnotesize{\texttt{0xcc7c...e78e}}} &
         \cellcolor{blue!20!white} \href{https://sepolia.arbiscan.io/tx/0x9858cff8322987c07504e150cbef94112e025acb0de4a37f9df045d3c7f38841}{\footnotesize{\texttt{0x9858...8841}}} & 
         \cellcolor{blue!20!white} \href{https://sepolia-optimism.etherscan.io/tx/0x9fea892e2b743206317527811e0e35b86e8a732636c1682cb0ac73ef7b68d918}{\footnotesize{\texttt{0x9fea...d918}}}  & 
         \cellcolor{blue!20!white} \href{https://sepolia-era.zksync.network/tx/0xc7665bcc19c8220ac43f651c51eb31238a95f0f9198c8e749d5114af36b05ab6}{\footnotesize{\texttt{0xc766...5ab6}}} &
         \cellcolor{blue!20!white} \href{https://sepolia.arbiscan.io/tx/0xed70e6167004cb6a13cec3a63b308b6cfb05be324ff57cf33fef6ce94cb62854}{\footnotesize{\texttt{0xed70...2854}}} & 
         \cellcolor{blue!20!white} \href{https://sepolia-optimism.etherscan.io/tx/0x69cdf5b56453dc9596b410dd6d2bb0fbdf217a265171639f86d10bc69b8274e8}{\footnotesize{\texttt{0x69cd...74e8}}} & 
         \cellcolor{blue!20!white} \href{https://sepolia-era.zksync.network/tx/0x70bf2a717ff3b6c91d9cd9c6553a33fbd1c17fcd4b0b8b65735e8ccbbdf242ec}{\footnotesize{\texttt{0x70bf...42ec}}}
         \\ \cline{2-11}
         \multirow{2}{*}{$T_{A_2}$} 
         & L1 & 
         \cellcolor{red!20!white} \href{https://sepolia.etherscan.io/tx/0x35e8dbdd0710ca87a48e53b31ae1804c40dd79b04a2c394558b7ba74b25843e0}{\footnotesize{\texttt{0x35e8...43e0}}} & \cellcolor{red!20!white} \href{https://sepolia.etherscan.io/tx/0x0578f49b95ffc708f99cfb5212a8ec022896341ddb0c0cf8b3e381cf3fd35e73}{\footnotesize{\texttt{0x0578...5e73}}} & \cellcolor{red!20!white} \href{https://sepolia.etherscan.io/tx/0x6e0a5a7af9cc5d75457c0bd3c4a9d6860c8d31cde206ff1081fe4b27ab3abeb3}{\footnotesize{\texttt{0x6e0a...beb3}}} 
         & - & - & - 
         & - & - & - \\
         & L2 & 
         \cellcolor{red!20!white} \href{https://sepolia.arbiscan.io/tx/0x31fa8e5e4fc08ad5a0d74cb81f2d2a9278e59d44c3e1075d99da73ea45645962}{\footnotesize{\texttt{0x31fa...5962}}} & 
         \cellcolor{red!20!white} \href{https://sepolia-optimism.etherscan.io/tx/0x595ab89c67914b9d5993ee932261a1960dd0506aebeee38418d06358fd943c0b}{\footnotesize{\texttt{0x595a...3c0b}}} & 
         \cellcolor{red!20!white} \href{https://sepolia-era.zksync.network/tx/0xd8ae6a84bed853e52b58b0c7e393283f2a1743f8c66cbe69dc74af65790c851b}{\footnotesize{\texttt{0xd8ae...851b}}} &
         \cellcolor{red!20!white} \href{https://sepolia.arbiscan.io/tx/0xa6c62d7261638c0129e0528f49776222a537df5a77a2a016f472743c7d96ae00}{\footnotesize{\texttt{0xa6c6...ae00}}} & \cellcolor{red!20!white} \href{https://sepolia-optimism.etherscan.io/tx/0x8ebf27abb3dd978b83af659812722b5e708264f5955e27def290febd23560394}{\footnotesize{\texttt{0x8ebf...0394}}} & \cellcolor{red!20!white} \href{https://sepolia-era.zksync.network/tx/0x080116647334dcc3302433f9aa8b4bf900747b310669013274e2ad83a17fd346}{\footnotesize{\texttt{0x0801...d346}}} &
         \cellcolor{red!20!white} \href{https://sepolia.arbiscan.io/tx/0xae2d773d0569360684201c315f721a70c289ef2c17916d046e2f4a59eb2e039d}{\footnotesize{\texttt{0xae2d...039d}}} & 
         \cellcolor{red!20!white} \href{https://sepolia-optimism.etherscan.io/tx/0xf24d84d05ca89f249fa07688a04b9f5b84abb73d32339b327458494b15cf359c}{\footnotesize{\texttt{0xf24d...359c}}} & 
         \cellcolor{red!20!white} \href{https://sepolia-era.zksync.network/tx/0x86e619f95deb062bb64632e883914cb93834e9239d15dc1d6d29ea8dbe902023}{\footnotesize{\texttt{0x86e6...2023}}} \\
        \bottomrule 
    \end{tabular}
    \end{adjustbox}
    \caption{Sepolia testnet transactions for each cross-layer sandwich attack strategy across Arbitrum, Optimism, and zkSync.}
    \label{tab:testnet_sandwiching}
\end{table*}

\section{List of Used Events}
\label{sec:events}

\tableautorefname{} \ref{tab:events} provides an overview of all the events that have been used in this paper to extract past information from the individual chains. Event topic hashes are grouped by individual categories.

\section{Detailed MEV Extraction Overview}
\label{sec:appendix_c}

\Cref{tab:arb_lib_count} provides the total number of arbitrages, liquidations, and sandwiches extracted across Ethereum, Arbitrum, Optimism, an zkSync for our measurement period of nearly 3 years. It also provides additional granularity on the distribution of different MEV types across different DeFi protocols.

\section{List of Identified Potential Victims}
\label{sec:victims}

\tableautorefname{} \ref{tab:victims} provides an overview of all L1 to L2 transactions across Arbitrum and Optimism that we identified as potential victims for our cross-layer sandwich attacks.

\section{Cross-Layer Sandwich Profits}
\label{sec:appendix_d}

\tableautorefname{} \ref{tab:sandwich_profitability} provides a detailed overview on the profitability of our cross-layer sandwich attack strategies $S_1$, $S_2$, and $S_3$ across various attacker budgets ranging from 1K USD capital to an infinite capital.

\section{Testnet Transactions}\label{sec:appendix_e}

\Cref{tab:testnet_sandwiching} provides an overview of successfully deployed transactions on the Sepolia testnet for the three attack strategies across Arbitrum, Optimism, and zkSync. Attacker transactions are highlighted in red, whereas victim transactions are highlighted in blue. 
These transactions serve a proof-of-concept to show that attackers could easily write scripts to automatically perform cross-layer sandwich attacks in real-time.

\end{document}